\documentclass[11pt,preprint]{aastex}

\def\gax{\mathrel{\raise.3ex\hbox{$>$}\mkern-14mu\lower0.6ex\hbox{$\sim$}}}
\def\lax{\mathrel{\raise.3ex\hbox{$<$}\mkern-14mu\lower0.6ex\hbox{$\sim$}}}
\def\gtorder{\mathrel{\raise.3ex\hbox{$>$}\mkern-14mu
             \lower0.6ex\hbox{$\sim$}}}
\def\ltorder{\mathrel{\raise.3ex\hbox{$<$}\mkern-14mu
             \lower0.6ex\hbox{$\sim$}}}

\begin{document}
\title{A Study of Cepheids in M81 with the Large Binocular Telescope \\
(Efficiently Calibrated with HST)}

\author{J.R. Gerke$^{1}$, C.S. Kochanek$^{1,2}$, J.L. Prieto$^{3,4}$, K.Z. Stanek$^{1,2}$, L. M. Macri$^{5}$}
 
\altaffiltext{1}{Department of Astronomy, The Ohio State University, 140 West 18th Avenue, Columbus OH 43210.}
\altaffiltext{2}{Center for Cosmology and AstroParticle Physics, The Ohio State University, 191 W. Woodruff Avenue, Columbus OH 43210.}
\altaffiltext{3}{Carnegie Observatories, Pasadena, CA, 91101.}
\altaffiltext{4}{Hubble and Carnegie-Princeton Fellow}
\altaffiltext{5}{George P. and Cynthia Woods Mitchell Institute for Fundamental Physics and Astronomy, Department of Physics \& Astronomy, Texas A\&M University, 4242 TAMU, College Station, TX 77843-4242.}

\begin{abstract}
We identify and phase a sample of 107 Cepheids with $10<P<100$~d in M81 using the Large Binocular Telescope and calibrate their B, V and I mean magnitudes with archival {\it Hubble Space Telescope} ({\it HST}) data.  The use of a ground-based telescope to identify and phase the Cepheids and {\it HST} only for the final calibration reduces the demand on this highly oversubscribed spacecraft by nearly an order of magnitude and yields Period-Luminosity (PL) relations with dispersions comparable to the best LMC samples.
We fit the sample using the OGLE-II LMC PL relations and are unable to find a self-consistent distance for different band combinations or radial locations within M81.  We can do so after adding a radial dependence to the PL zero point that corresponds to a luminosity dependence on metallicity of $\gamma_{\mu}=-0.56\pm0.36$~mag dex$^{-1}$. We find marginal evidence for a shift in color as a function of metallicity, distinguishable from the effects of extinction, of $\gamma_2=+0.07\pm0.03$~mag dex$^{-1}$. We find a distance modulus for M81, relative to the LMC, of $\mu_{M81}-\mu_{LMC}= 9.39\pm0.14$~mag, including uncertainties due to the metallicity corrections. This corresponds to a distance to M81 of $3.6\pm0.2$ Mpc, assuming a LMC distance modulus of 18.41 mag.
We carry out a joint analysis of M81 and NGC~4258 Cepheids and simultaneously solve for the distance of M81 relative to NGC~4258 and the metallicity corrections. Given the current data, the uncertainties of such joint fits are dominated by the relative metallicities and the abundance gradients rather than by measurement errors of the Cepheid magnitudes or colors. We find $\mu_{M81}-\mu_{LMC}=9.40_{-0.11}^{+0.15}$~mag, $\mu_{N4258}-\mu_{LMC}=11.08_{-0.17}^{+0.21}$~mag and $\mu_{N4258}-\mu_{M81}=1.68\pm0.08$~mag and metallicity effects on luminosity and color of $\gamma_\mu=-0.62_{-0.35}^{+0.31}$~mag dex$^{-1}$ and $\gamma_2=0.01\pm0.01$~mag dex$^{-1}$. Quantitative analyses of Cepheid distances must take into account both the metallicity dependencies of the Cepheids and the uncertainties in the abundance estimates.

\end{abstract}


\section{Introduction}
\label{sec:introduction}

Cepheid variables have long been one of the key links in the distance ladder, both for measuring distances to nearby galaxies and for estimates of the Hubble Constant. Improving on the accuracy of the estimates by the first generation of {\it HST}-based projects (\citealt{freedman2001}; \citealt{sandage2006}) requires improved distances to at least one calibrating galaxy and better characterizations of the systematic uncertainties, principally the effects of metallicity and blending (e.g., \citealt{kochanek1997}; \citealt{kenni1998}; \citealt{groen2004}; \citealt{sakai2004}; \citealt{macri2006}; \citealt{scowcroft2009}; \citealt{riess2009}; \citealt{bono2010}; \citealt{shappee2010} and  \citealt{stanek1999,moche2000,ferra2000,evans2008}).  Considerable progress has been made on the first point, principally by using NGC 4258 with its maser distance as the distance calibrator \citep{herrn1999,humph2008}.  However, a better characterization of systematic effects in the Cepheid Period Luminosity (PL) relation has not had such a clear resolution.

The effect of metallicity on the Cepheid PL relations is one of the most hotly-debated aspects of the Cepheid distance scale. Cepheid abundances are assumed to correlate with the gas-phase oxygen ($[O/H]$) abundance and gradient in the disk of their host galaxies.  The absolute abundances are unimportant, only accurate relative abundances are needed.  These relative abundances are generally estimated from spectroscopy of \ion{H}{2}~regions, usually using the $R_{23}$ method based on the line flux ratio $([O_{II}]+[O_{III}])/H\beta$.  Empirical estimates of the effect generally show metal-rich Cepheids to be brighter than metal-poor ones (\citealt{gould1994}; \citealt{sasselov1997}; \citealt{kochanek1997}; \citealt{kenni1998}; \citealt{groen2004}; \citealt{sakai2004}; \citealt{macri2006}; \citealt{scowcroft2009}), although some find the opposite (\citealt{roman2008}).  The {\it HST} Key Project on the Extragalactic Distance Scale \citep{freedman2001} obtained data in only two bands ($V$ and $I$), and estimated extinction from the $V-I$ color. They adopted a metallicity correction of $\gamma_{VI}=-0.2\pm0.2$ mag dex$^{-1}$, largely based on the apparent difference in the Cepheid distance between an inner, high metallicity and an outer, low metallicity field of M101 \citep{kenni1998}. However, \citet{shappee2010} recently found a metallicity correction for M101 of $\gamma=-0.83\pm0.21$ mag dex$^{-1}$ based on a larger Cepheid sample and a revised metallicity gradient (\citealt{bresolin2007}, also see \citealt{bresolin2011a}, \citealt{bresolin2011b}).  In contrast, theoretical predictions of the effect at optical wavelengths predict a negligible effect of $\gamma_{VI}\sim+0.03$ mag dex$^{-1}$ for $W_{VI}$ \citep{bono2008}.  Limited studies in the $H$ band have found a smaller effect of $\gamma=-0.23\pm0.17$~mag dex$^{-1}$ \citep{riess2009} and $\gamma=-0.10\pm0.09$~mag dex$^{-1}$ \citep{riess2011}, in better agreement with theoretical expectations that Cepheid magnitudes should be less affected by metallicity at near-infrared wavelengths \citep{marconi2005}. An accurate characterization of the metallicity dependence of the Cepheid PL relation requires data in multiple bands for large numbers of Cepheids at a common distance in order to separate the effects of extinction from that of metallicity and to exploit abundance gradients to test for their effects. As we will eventually conclude, these relative (but not absolute) metallicities need to both be better determined and have their uncertainties fully included in the analysis of the Cepheids. 

Blending is the second major systematic uncertainty beyond distance zeropoint errors.  Blending occurs when a Cepheid has a close visual companion that contaminates its PSF, causing the variable to appear artificially brighter and leading to an underestimate of the distance.  This can be due to true binary companions \citep{evans2008}, stars correlated with the Cepheid, or chance projections. There is no consensus on the degree to which blending affects distance estimates \citep{stanek1999,moche2000,ferra2000}, and it is clear that the effects of blending need to be studied further.  For example, the null result of \citet{ferra2000} ignored the strong spatial correlations of luminous stars \citep[see][]{harris1999}, probably leading to an underestimate of the effect compared to the empirical study by \citet{moche2000}.  As with metallicity effects, however, blending effects can be controlled using multiple bands because the vast majority of blended stars must be either bluer or redder than the rare yellow Cepheids, and thus modify their extinction-corrected colors.   

Beyond the Local Group, astronomers have relied on space-based observations from {\it HST}, with its superior resolution, to identify Cepheids in other galaxies.  While very successful, these studies suffer from two major limitations.  First, only small fields are surveyed, so the samples in any galaxy tend to be smaller and/or biased towards fainter, shorter period Cepheids.  Second, the high cost of the monitoring needed to recognize the Cepheids and determine their periods has meant that data is obtained for the smallest possible number of epochs and in very few bands, limiting the ability to search for and study systematic problems.  
For nearby galaxies, there is no reason to do the expensive monitoring using {\it HST} or any other space-based observatory.  Long-period Cepheids can be identified and phased relatively easily from the ground, even at 10 Mpc. The key technology is difference imaging (e.g. \citealt{alard2000}), which allows for efficient detection of variable sources even in crowded fields.  Its power was illustrated by \cite{bonanos2003}: where traditional photometry identified only 12 Cepheids in VLT observations of M83 (\citealt{thim2003}), difference imaging successfully identified 112. Once Cepheids are identified, only a single epoch of space-based data is theoretically needed to calibrate the differential light curves found with image subtraction.  If there are significant color terms to the absolute calibration, the calibrating data needs to be obtained at a common epoch.

Here we present results from monitoring the entire disk of M81 using the twin 8.4-m Large Binocular Telescope \citep{hill2006}. We have identified 140 Cepheids in this galaxy to date, using image subtraction techniques.  After phasing the light curves and fitting them to templates, we were able to match and calibrate 126 of these variables using archival HST/ACS images in the B, V and I bands based on the catalogs from \citet{dalcanton2009}.  After applying additional cuts based on data quality and physical parameters, we have a final sample of 107 Cepheids compared to the 17 in the final {\it HST} Key Project sample.  With three bands and two colors, we can estimate the distance to M81, the extinction to each individual Cepheid, and still have radial positional gradients and one more color to search for physical effects due to metallicity and blending.  We give details of the observations and the data reduction in \S2.  \S3 explains our approach to light curve calibration and \S4 presents the initial PL relations and our exploration of systematic problems. We look into the physical effects of metallicity (radial position) on the PL relations in \S5 and expand on this by jointly analyzing our M81 sample with that from \citet{macri2006} in the maser calibrated \citep{herrn1999} galaxy NGC~4258 in \S6.  We discuss the results, their implications and outline our future plans in \S7. 

\section{Observations and Data Reduction}

The galaxy M81 is being monitored as part of a ground-based variability survey of nearby galaxies \citep{kochanek2008} with the twin 8.4-m Large Binocular Telescope (LBT). The survey is monitoring 25 galaxies within 10 Mpc with high star formation rates to look for failed supernovae and to study supernova progenitors. The companion tidal dwarf galaxy Holmberg IX also falls in the field of view.  \citet{prieto2008} discovered a massive eclipsing binary in Holmberg IX using the preliminary results of the project.  The observations discussed here were taken with the LBC-Blue camera \citep{giallongo2008} on 34 nights between 2007 January and 2008 May.  Some nights were subdivided to yield a total of 50 epochs in the V band.  The cadence and depth of the observations allowed detection of Cepheids with periods ranging from 10 to 100 days.  Multiple images were taken during each observation and these images were then combined using a sigma-clipped average.  The exposure times were 60 and 120 seconds for the 2007 and 2008 observations, respectively.  

We used the IRAF\footnote{IRAF is distributed by the National Optical Astronomy Observatories, which are operated by the Association of Universities for Research in Astronomy, Inc., under cooperative agreement with the National Science Foundation} MSCRED package to perform the basic reduction of the mosaic images: overscan correction, bias subtraction and flat fielding using twilight skyflats usually obtained the same night as the science images.  An initial astrometric solution was found using astrometry.net \citep{lang2010}.  Next, the images were processed using the ISIS image subtraction package \citep{alard2000,lupton1998}.  Image subtraction works by matching a reference image in flux and PSF structure to each epoch and subtracting it to leave only the time variable flux of the sources. The reference images were created from the median of the 17 highest-resolution epochs.  The coordinates reported in this work are based on astrometric solutions for these reference images, found using WCSTools \citep{mink2002} with the Sloan Digital Sky Survey catalog as a reference.  The coordinates have errors $\sim 0\farcs1$.  Variable sources were identified in the ``absdiff'' image found by convolving the subtracted images with a $\sigma=2$ pixel Gaussian, summing their absolute values and then median filtering the background. We used Sextractor \citep{bertin1996} to identify the variable source positions in the ``absdiff'' image and then used ISIS to construct light curves for all sources found in M81 using the differential flux for each observation. Each light curve was analyzed following \citet{schwarz1989} (Analysis of Variance) to determine the likelihood of being a variable source and to estimate its period.  We examined all variable light curves by eye, and flagged as Cepheids the periodic sources with the characteristic ``sawtooth'' light curve shape.

The entire disk of M81 was imaged using HST/ACS and the F435W and F606W filters (roughly equivalent to the $B$ and $V$ bands) by program 10584 (PI: Zezas) and the F814W filter (roughly equivalent to the $I$ band) by program 10250 (PI: Huchra). The HST images were acquired between 2004 September and 2006 December, but we will have no trouble phasing the LBT light curves to these earlier dates thanks to the long time span of the ground-based data (see \S3). ACS was used to observe M81 at 29 tiled positions with the F435W and F606W filters and at 24 tiled positions with the F814W filter, roughly covering the same area as our LBT survey.  We used the M81 photometric catalogs created by the ANGST collaboration \citep{dalcanton2009}.  Their B- and V-band catalogs are publicly available through the ANGST website\footnote{http://www.nearbygalaxies.org}. The I-band catalog is not currently public and was generously provided by the ANGST team (B. F. Williams, private communication). These catalogs contain the standard ANGST data products and were generated using DOLPHOT \citep{dolphin2000}.  The I-band magnitudes were transformed from the HST F814W bandpass to the standard Kron-Cousins system using the prescription from \citet{sirianni2005}. Thus, all HST magnitudes used in this paper are calibrated to the standard Johnson-Cousins B, V and I system.

We used the complete ANGST catalogs, without cuts on signal-to-noise ratio (SNR), sharpness or crowding, but kept track of these quality flags.  We made an initial match to the Cepheids found with LBT using the objects in the ANGST catalog within $1\farcs0$.  This initial matching criteria was large because the ANGST coordinates are based on the {\it HST} pointing astrometry and can have absolute errors of $1\farcs0$ or more.  We then matched the brightest 15 stars in an HST field to the LBT image and found the average coordinate shift needed to align the HST image with the LBT data.  We were then able to narrow the matching criteria to $\sim 0\farcs06$.  All matches were checked by eye to ensure accuracy.

\section{Calibration Procedure}

We do not want to simply treat the HST data as random phase observations and use them as ``mean magnitudes'' for the PL relation.  Instead we want to match the HST observations to the LBT light curves and accurately estimate true mean magnitudes. We used the LBT V band light curves to determine the period of each Cepheid, which was then applied to all three bands to phase the HST points and determine the magnitude calibrations. The V and I bands were fit with templates from \citet{stetson1996}.  The B band template was modeled by scaling the amplitude of the V band template by a factor of $1.51 \pm 0.20$ based on the mean amplitude ratio of the 281 NGC 4258 Cepheids found in \citet{macri2006}.  The \citet{stetson1996} templates are optimized for Cepheids with periods from 10 to 100 days and we restricted our sample to this range, although in practice we found few objects outside of it.  The goodness-of-fit to the V band light curve was used as a quality cut.  We removed the 25\% of the sources with the largest $\chi^2$ --- any objects that did not have a well defined period, had large errors in photometry or were otherwise poorly fit by the template were dropped.  This resulted in a sample of 126 Cepheids and removed about 45 sources from our original list of candidates.  While many of these rejected sources may be Cepheids, the light curves were not as clean as the rest and we can afford to be conservative given this large sample.  One of the most frequent causes of poor light curve quality was noise from a nearby bright star.  As we determine the period and phasing of a Cepheid, we simultaneously solve for the other parameters needed to characterize the Cepheid and transform the differential measurements into apparent magnitudes. 

Because of the time span of our light curves, we can accurately estimate the Cepheid phase at the time of the HST observations.  The problem is determining the amplitude. The magnitude of a Cepheid at phase $\phi$ is
\begin{equation}
M=\langle M \rangle+A T(\phi)
\end{equation}
where $\langle M \rangle$ is the mean magnitude, T($\phi$) is the template at phase $\phi$, and A is the amplitude. Thus, the mean magnitude is related to the HST calibration magnitude $M_{HST}$ by 
\begin{equation}
M_{HST}=\langle M \rangle+A T(\phi_{HST}),
\end{equation}
where $\phi_{HST}$ is the phase of the HST observation.  We estimate the amplitude by fitting the difference imaging light curves with counts $\Delta$C($\phi_i$) at phase $\phi_i$, 
\begin{equation}
\Delta C(\phi_i)=10^{-0.4[\langle M \rangle+A T(\phi_i)-Z]}-C_0
\end{equation}
where $C_0$ is the (unknown) counts of the Cepheid in the reference image and Z is the photometric zeropoint of the reference image.  As a reminder, if we could accurately determine $C_0$ from the crowded LBT observations, we would have no need for HST calibrations.  We simultaneously fit the HST calibration point and the differential light curve to determine $\langle M \rangle$, A, and $C_0$.  The uncertainties were then calculated using Monte Carlo Markov Chains.

To determine the zeropoint for each chip of the LBC camera in the reference images, we obtained photometry of the V band reference images using DAOPHOT \citep{stetson1987} and compared it to the SDSS catalog photometry, transformed to Johnson-Cousins magnitudes \citep{fukugita1996}.  We removed any sources flagged as possible variables and restricted the comparison to $V<21$~mag. Care was taken to ensure the entire field was covered and that crowded regions were avoided.  This resulted in about 60 stars per LBC chip with which to separately calculate the zeropoints. The accuracies ($\pm0.07$ mag) of these zeropoints only indirectly affect the estimate of $\langle M \rangle$ through their effect on our estimate of the amplitude, as we explore further in \S4.

If we simply used random phase estimates for the mean magnitude, its uncertainty would be

\begin{equation}
\sigma_{\langle M \rangle}^2=\sigma_{HST}^2+A^2 \langle T^2 \rangle
\end{equation}

\noindent where $\sigma_{HST} \simeq 0.013$~mag is the typical uncertainty in the V-band HST photometry and $\langle T^2 \rangle ^{1/2}$ is the rms average of the template.  Our template models have $\langle T^2 \rangle ^{1/2}$=1.14, 0.75, and 0.44 for the B, V, and I bands, respectively, and the typical amplitude is $A=0.36 \pm 0.03$~mag.  Note that $\langle T \rangle=0$ by definition.  Thus, random phase calibrations would have typical uncertainties in  $\langle M \rangle$ of order $\sigma_{\langle M \rangle}$= 0.25, 0.21, and 0.15 mag for the B, V, and I bands respectively.  When we use the ground based light curves to estimate the amplitude and phase, the uncertainty is 
\begin{equation}
\sigma_{\langle M \rangle}^2=\sigma_{HST}^2+\sigma^2_A  T^2(\phi_{HST})+\sigma_{\phi}^2A^2\left(\frac{\delta T(\phi_{HST})}{\delta \phi}\right)^2
\end{equation}
where $\sigma_A$ is the uncertainty in the amplitude and $\sigma_{\phi}$ is the uncertainty in the phase of the HST observations.  

While the HST data were obtained $\Delta T=1-3$ years before the LBT data, the $\Delta T_{LBT} = 1.3$ year time span of the LBT observations means we can accurately determine the phase at the time of the HST observations essentially by time reversal. Suppose that the allowed phase error for a well-fit light curve over the time span $\Delta T_{LBT}$ is $\sigma_{\phi_{LBT}}$.  A small phase error only grows linearly with time, so 

\begin{equation}
\sigma_{\phi_{HST}}=\sigma_{\phi_{LBT}} \left(\frac{\Delta T}{\Delta T_{LBT}}\right),
\end{equation}

\noindent and the resulting uncertainties must be small unless the phasing of the LBT data is poor, which would have led to its rejection as described above. We also checked for errors in period and phase, examining the residuals from the PL as a function of the number of Cepheid periods between the HST observation and the start of the LBT data and found no trends.  We also looked at the residuals as a function of the phase of the HST data and again found no trend.  Errors in the period, which would translate to errors in the phase of the HST point, appear to be insignificant. Given that the phase errors are unimportant, we will do better than random phases if the fractional error in our amplitude estimate is smaller than

\begin{equation}
\frac{\sigma_A}{A} < \left \vert \frac{\langle T^2 \rangle ^{1/2}}{T(\phi_{HST})}\right \vert,  
\end{equation}

\noindent where the minimum value of $ \left \vert \langle T^2 \rangle ^{1/2}/T(\phi_{HST})\right \vert$ =0.57, 0.57 and 0.62 for the B,V and I bands. These criteria are easily satisfied. See Figure 1 for representative examples of the light curve fits.

Thanks to the overlap of the ACS fields, two HST observations were available for 13 of the Cepheids and we used them to check our calibrations. While we have two points in both B and V for these overlapping Cepheids, only the V band affects the light curve fit because the amplitudes and periods are determined using the V band and then are applied to the other bands. We solved for the calibration of these Cepheids twice, once for each observation. One of the fits to Cepheid M81C~095616.57+685615.1 yielded a non-physical $\bv$ color, and there was no light curve fit consistent with both calibration points.  The non-physical color is likely due to an error in the HST V band magnitude, perhaps caused by a cosmic ray. The calibration point that gave the non-physical color was rejected, leaving 12 Cepheids with 2 calibration points. For these 12 Cepheids, we found little difference in the individual periods and phases of the Cepheids between the two fits with different HST observations.  We find average differences of $\langle \Delta M_V \rangle$=$\langle \Delta M_B \rangle$=$0.09\pm0.02$~mag with $\sigma_V=0.24$~mag and $\sigma_B=0.34$~mag.  If we remove the three largest outliers from each band, the average differences drop to $\langle \Delta M_V \rangle$=$0.01\pm0.01$~mag and $\langle \Delta M_B \rangle$=$0.01\pm0.02$~mag with $\sigma_V=0.11$~mag and $\sigma_B=0.16$~mag.  Two of the three largest outlier Cepheids were common between the B and the V bands.  The B and V band calibration observations were taken at the same epoch.
% The average difference in $\langle M_V \rangle$ is $0.09\pm0.02$~mag with a scatter of $\sigma=0.24$~mag, although if we remove 3 outliers, the average difference drops to $0.01\pm0.01$~mag with $\sigma=0.11$~mag.

We investigated the outliers to understand the change in the mean magnitudes with different calibration points. We first looked at the photometry quality indicators from the HST data. None of the Cepheids in this sub-sample are outliers in terms of photometric quality.  Moreover, any difference in the quality of the two calibration points did not correlate with difference in mean magnitude. The remaining sources of uncertainty are related to the light curve template fit, specifically the amplitude A and the unknown counts $C_0$ in the reference image.  Since only the V band data determines the amplitude, we will now focus on the V band.  We found that the Cepheids with the largest $\Delta\langle M_V \rangle$ also had the largest difference in their amplitude estimates. To investigate this trend further, we fit the light curve while keeping the amplitude fixed and found ---as expected--- that the shifts in $\Delta\langle M_V \rangle$ were due to shifts in amplitude.  Thus, the largest source of uncertainty in determining the mean magnitudes from the differential light curves with a single HST calibration point is, as expected, the amplitude of the light curve.  However, for our present data, such errors should be uncorrelated with distance, environment and metallicity, and the large number of Cepheids will reduce the effect of any amplitude errors.  For the overlapping objects, we determined the light curve fits using both V-band HST calibration points, giving a single estimate for $\langle M_V \rangle$.  We also averaged the values from the two B-band calibrations, which typically agreed at the 1$\sigma$ level.  Table~\ref{tab:ceph} lists the coordinates, periods, and calibrated phase-averaged mean magnitudes in the B, V and I bands for the 126 Cepheid variables in our sample.  Some of these Cepheids are flagged for being calibrated using multiple HST epochs from ANGST or from \citet{freedman1994}, as we discuss below.

We can also verify our results using Cepheids in common with other {\it HST} surveys. Unfortunately there was no overlap between our LBT survey area and the ANGST study of Cepheids in the outer disk of M81 by \citet{mccommas2009}.  We do overlap the WFPC survey of \citet[][hereafter F94]{freedman1994}. For this comparison it should be noted that the F94 data were obtained with the first-generation WFPC instrument which had no aberration correction.  Our final sample contains 11 of the 31 F94 Cepheids, where most of the Cepheids we missed are in highly extincted areas.  Table~\ref{tab:ceph} flags the Cepheids found by both surveys. We compared the periods, the amplitudes and mean magnitudes of these Cepheids.  The average fractional difference in period is $\langle \Delta P/P \rangle$=0.03, where the F94 periods are on average shorter.  Three of the Cepheids in common had period differences of more than 2 days. The most dramatic period difference was found for Cepheid C26 in F94 or M81C~095610.62+690732.7 in our study, where the F94 period is 54.8~days and ours is 64.54 days. Figure \ref{fig:KPceph} shows both the F94 data and our LBT data, phased to both the F94 period and the LBT period.  The F94 light curves seem well phased for both periods, despite the 10 day difference, while the LBT data does not phase correctly at the F94 period.  The long baseline and number of epochs of the LBT data enables us to more accurately determine the period. Excluding this object, the average fractional period difference is $\langle \Delta P/P \rangle$=0.02. 

%here
We next compared the V band template amplitude parameters of the light curves.  We first fit the F94 light curves using the template models and compared the estimated template amplitudes.  We used our period estimate for the discrepant Cepheid (C26/M81C~095610.62+690732.7).  There were only ~6 observations in the I band by F94, so the I band data were fit with templates using the light curve parameters determined by the V band data.  We found an average template amplitude parameter difference of $\langle \Delta A \rangle =-0.08 \pm0.02$ or a fractional shift of  $\langle \Delta A/A \rangle =-0.26 \pm0.04$, where the LBT amplitudes are generally smaller.  The conversion from the template parameter shift $\Delta A$ to the change in the mean magnitude depends upon the specific Cepheid, but the average factor to convert $\Delta A$ to magnitudes is about 2.4.  The Cepheid with the largest difference is C9/M81C~095502.67+690954.4.  This Cepheid happens to be one for which we also have two ACS calibration points.  The fit to the individual calibration point that had the template amplitude parameter closest to that of the F94 fit still had a magnitude difference of 0.14 and the fit with both calibration points had $\Delta A=0.19$, so it remains an outlier between our results and F94 even when we use both ACS calibration points.  If we remove this Cepheid, the average template amplitude parameter difference is $\langle \Delta A \rangle =-0.07 \pm0.02$.  We examine the consequences of this as a possible bias in \S4.

Finally, we investigated how our mean V and I band magnitudes compared to those from F94.  We compared our mean magnitudes to both the mean V and I magnitudes originally reported by F94 and to the mean magnitudes we calculated by fitting the Stetson templates to the F94 data.  \citet{freedman1994} reported the mean magnitudes found by averaging all data points with uncertainties below 0.3 mag.  Since the errors depend on magnitude and the magnitude depends on phase, this can be a dangerous practice because it biases the mean magnitudes to be brighter as photometric errors increase.  The average difference between the original F94 magnitudes, found through averaging observations with uncertainties below 0.3 mag, and the mean magnitudes we find with a template fit to the F94 data is 0.05$\pm0.01$~mag with a dispersion of $\sigma=0.08$~mag for the V band and 0.03$\pm0.02$~mag with a dispersion of $\sigma=0.10$~mag for the I band, where the template estimates are fainter.  This is in the same direction as the bias expected from rejecting higher uncertainty (fainter) points. Our mean magnitudes based on the HST-calibrated LBT data differ from the original F94 mean magnitudes by an average of $0.13\pm0.02$~mag with a dispersion of $\sigma=0.18$~mag  in the V band and by $0.12\pm0.02$~mag with a dispersion of $\sigma=0.24$~mag in the I band.  Our mean magnitudes are generally dimmer.

The Cepheid which had the largest difference in period also had the largest difference in V band mean magnitude, at 0.42~mag (see Figure \ref{fig:KPceph}).  The Cepheid with the largest difference in the I band had the largest difference in the amplitude parameter.  We carried out a fit to the period-luminosity relation without metallicity corrections to match the F94 procedure and found that our calibrations of these Cepheids (C26 and C9 in F94 or M81C~095610.62+690732.7 and M81C~095502.67+690954.4) have PL residuals smaller than $1\sigma$ in all three bands. This would not be true if we adjusted the mean magnitude by the respective differences of 0.42~mag and 0.49~mag. \citet{freedman1994} noted that C26 was elongated in their images and had a nearby companion and that C9 had a faint nearby companion, perhaps explaining the brighter magnitude in their analysis. However, the differences in mean magnitude do not generally correlate with the notes on the environment of each Cepheid from F94, with the ANGST error flags for photometry, or with light curve quality.  Removing the outlier reduces the average difference in the mean magnitudes to $0.10\pm0.01$ mag in the V band and $0.07\pm0.01$ mag in the I band.  

When we compare our HST-calibrated LBT mean magnitudes to those based on the template fits to the F94 data (rather than the published F94 means) we find the average difference decreases to $0.08\pm0.02$ mag with a dispersion of $\sigma=0.17$~mag in the V band and $0.09\pm0.02$ mag with a dispersion of $\sigma=0.17$~mag in the I band.  C26/M81C~095610.62+690732.7 was again the most discrepant in the V band, with a magnitude difference of 0.35 mag.  The largest difference in the I band was 0.27 mag for C29/M81C~095613.56+690620.0.  Removing these outliers gives an average mean magnitude difference of $0.05\pm0.02$ mag, with a dispersion of $\sigma=0.16$~mag in the V band and $0.07\pm0.02$ mag, with a dispersion of $\sigma=0.18$~mag in the I band. 
The average V-band (I-band) magnitude difference is unchanged (shifts
by $1\sigma$) compared to those seen betweeen the original F94 magnitudes and those from template fits to the F94 data.  We used the F94 data to update the calibrations of these 11 Cepheids.  This lead to an average change of 0.01~mag in our mean V band magnitudes and $-0.01$~mag in our I band magnitudes. In all, 22 of our Cepheids are calibrated using multiple HST epochs from F94 or ANGST.

\section{Period-Luminosity Relations}

\label{sec:pls}
The periods and mean magnitudes of the 126 Cepheids listed in Table~\ref{tab:ceph} were fit to the updated OGLE II extinction-corrected PL relations \citep{udalski1999}\footnote{ftp://sirius.astrouw.edu.pl/ogle/ogle2/var\_stars/lmc/cep/catalog/README.PL},

\begin{eqnarray}
\label{eqn:pls}
B(P)&=&17.368(31)-2.439(46)\log P \nonumber \\
V(P)&=&17.066(21)-2.779(31)\log P \\
I(P)&=&16.594(14)-2.979(21)\log P \nonumber \\
W_{VI}(P)&=&15.910(46)-3.269(68)\log P. \nonumber
\end{eqnarray}

We constrained the Cepheids to lie at a common distance modulus, $\mu$, and allowed for individual extinctions, $E_i$.  The extinctions were estimated assuming a \citet{cardelli1989} extinction law with $R_B$=4.28, $R_V=3.28$, and $R_I=1.94$ from Table 6 of \citet{schlegel1998}.  We also examined the ``reddening-free'' Wesenheit index $W_{VI}=V-R \times (V-I)$ with R=2.45 \citep{madore1982} for the V and I bands in order to compare with the \citet{freedman2001} and \citet{mccommas2009} results. The definition in Equation (8) of $W_{VI}$ is identical to that used by \citet{mccommas2009}.

We fit the BVI PL relations and estimated $\Delta \mu_{LMC} =\mu -\mu_{LMC}$, the distance modulus of M81 relative to the LMC, the extinction, $E_i$, of each Cepheid $i$, along with their uncertainties, through a global $\chi^2$ minimization of

\begin{eqnarray}
\label{eqn:plsb}
\left( 
\begin{array}{c}
 \langle B_i \rangle  \\
 \langle V_i \rangle \\
 \langle I_i \rangle \\
\end{array} \right)
=
\left( 
\begin{array}{c}
B(P_i) \\
V(P_i) \\
I(P_i) \\
\end{array} \right)
+
\Delta \mu_{LMC}
\left( 
\begin{array}{c}
1 \\
1 \\
1 \\
\end{array} \right)
+
E_i 
\left( 
\begin{array}{c}
R_B \\
R_V \\
R_I \\
\end{array} \right),
\end{eqnarray} 

\noindent where $\langle B_i \rangle$, $\langle V_i \rangle$, and $\langle I_i \rangle$ are the mean magnitudes of Cepheid $i$, $B(P_i)$, $V(P_i)$, and $I(P_i)$ are the magnitudes expected from the PL relations in Equation~\ref{eqn:pls}, and $E_i$ is the estimated extinction for Cepheid $i$. $E_i$ is an estimate of the absolute extinction, assuming the red-clump method used by \citet{udalski1999} to correct LMC Cepheids for extinction is accurate.

We will use two methods to report our errors.  The first method simply uses standard $\chi^2$ statistics.  Because there is intrinsic scatter about the PL relations that is not included in the photometric error estimates, we first rescale the fits so that $\chi^2/{\rm dof}$=1, where ``dof'' is the number of degrees of freedom in the fit.  We then estimate the $1\sigma$ (68.3\% confidence) uncertainty on one parameter as the range with $\Delta\chi^2=1$. The second method uses bootstrap resampling.  We carried out $10^4$ trials randomly resampling the Cepheids and report the 68.3\% confidence region for comparison with the results from the $\chi^2$ statistic.  When we report a value for the bootstrap approach, it corresponds to the median of the trials.  Both error estimates are reported in Tables~\ref{tab:distnoz} and \ref{tab:distz}, although in our later analyses including the metallicity uncertainties we discuss only bootstrap results. 

We carried out an initial analysis using all 126 Cepheids having acceptable light curve fits, and found a distance modulus of $\Delta \mu_{LMC}=9.21\pm0.02$~mag for M81 relative to the LMC. The errors in extinction and distance were rescaled by a factor of 5.1 to make $\chi^2/{\rm dof}=1$. The bootstrap re-sampling error estimates yielded somewhat larger uncertainties of $\pm0.06$~mag.  The dispersions about the PL relations are $\sigma= 0.178$, $0.113,$ and $0.189$~mag for the B, V, and I bands, respectively, which is already comparable to the scatter seen in the OGLE II Cepheids in the LMC.  We then used these results to select outliers and to look for any trends with data quality.  We began by examining the outliers in extinction and amplitude. We also looked for correlations of PL residuals with the ANGST photometric quality information.  The PL residual is defined as the extinction-corrected mean magnitude of the Cepheid minus the expected mean magnitude, given the best-fit global distance modulus and the individual extinction of the variable.

While we found no trends with extinction or amplitude, there were outliers in both distributions, as can be seen in Figure \ref{fig:amp}.  We considered $E(\bv)$$\ge$0.4 mag to be high extinction and $E(\bv)$$\le$$-0.1$ mag to be low extinction because the distribution drops off at these limits.  We allowed $E(\bv)$$\le$$-0.1$ since there will be scatter around zero extinction and we want to avoid a bias toward higher extinction. Only one Cepheid stood out in amplitude, with $\hbox{A} > 0.9$,  while the rest of the sample had $\hbox{A} < 0.8$.  No other properties of these Cepheids appeared unusual.  Their light curve fits and HST photometric quality flags were typical of the full sample.  The Cepheid period amplitude distributions are consistent with those for Galactic \citep{klag2011} and OGLE III LMC \citep{soszynski2008} Cepheids.

Next we looked for trends in the residuals with the ANGST photometric quality indicators. Crowding was the first photometric quality parameter we examined. The crowding parameter is in magnitudes and gives the change in brightness of a star measured before and after subtracting the nearby sources.  A high crowding parameter could indicate a higher probability of blending. \citet{gogarten2009} suggest that a cut of $({\rm crowd}_V + {\rm crowd}_I) \le 0.1$ is a very conservative choice and $({\rm crowd}_V + {\rm crowd}_I) \le 0.6$ is a solid limit for maintaining photometric quality while including cluster stars.  Figure \ref{fig:crowd} shows the distribution of the crowding parameter as a function of the PL residuals.  While there are no clear trends, there are a few Cepheids with crowding $>0.2$ in one band.  The sharpness parameter, which estimates how well a star was fit by the PSF, is another indicator for blending.  It is positive for a star that is sharper than the PSF and negative if the star is broader.  Again we find no trend of sharpness with the PL residuals.  The conservative constraint placed on sharpness in the ANGST catalogs is $({\rm sharp}_V + {\rm sharp}_I)^2 \le 0.075$.  All our Cepheids have $({\rm sharp}_B + {\rm sharp}_V + {\rm sharp}_I)^2 \le 0.03$, well within these conservative sharpness limits. 

Based on these considerations, we adopted the following four selection cuts on our sample. (1) We limited the extinctions to the range $-0.1\le E(\bv)\le0.4$, removing 5 Cepheids. (2) The amplitude was limited to $0.1\le A\le0.8$. This led to dropping the one Cepheid with a very high amplitude and two with very low amplitude. (3) We required the crowding parameter to satisfy ${\rm crowd} \le 0.2$ in all filters and $({\rm crowd}_B + {\rm crowd}_V + {\rm crowd}_I) \le 0.4$ in total.  These criteria cut two Cepheids.  (4) Our sample met a sharpness criteria of $({\rm sharp}_B + {\rm sharp}_V + {\rm sharp}_I)^2 \le 0.03$ with no cuts.  With these additional criteria, we have removed the two Ultra Long Period (ULP) Cepheids with P$>$80 days from our sample. While ULP Cepheids are valuable for extending the distance over which Cepheids are effective distance indicators, they may follow a different PL relation \citep{bird2009}. 

The trimmed sample of 117 Cepheids yields a distance modulus of M81 relative to the LMC of $\Delta \mu_{LMC} = 9.18\pm0.02$~mag. {This is 0.03 mag smaller but within the uncertainties of our initial estimate.  Bootstrap re-sampling gives $\Delta \mu_{LMC} = 9.18\pm0.05$~mag, which is also within the uncertainties of our initial estimate.  The dispersion of the PL residuals in each band decreased somewhat from the initial fits, but there were still several outliers in the relations.  We examined the distribution of average absolute values of the residuals and found a natural break at 0.22~mag, which we adopted as a cut. This removed 10 Cepheids from the sample. 

The final sample of 107 Cepheids gives a relative distance modulus of $\Delta \mu_{LMC} = 9.19\pm0.02$~mag, where we now need to rescale the errors by a smaller factor of 3.0 to make $\chi^2/{\rm dof}=1$. The distance modulus is 0.01 mag larger than the previous estimate (i.e., before removing outliers from the PL relations), but still consistent given the uncertainties. The bootstrap re-sampling estimates are consistent with the $\chi^2$ estimates, and now have comparable uncertainties, with $\Delta \mu_{LMC} = 9.19\pm0.03$~mag, as might be expected if outliers drove the earlier differences between the two error estimates.  We also determined the distance using the random-phase HST data, and found $\Delta \mu_{LMC} = 9.22\pm0.03$~mag or bootstrap results of $\Delta \mu_{LMC} = 9.22\pm0.06$~mag.  Figure \ref{fig:PL3rand} shows the random phase PL relations for the B, V, and I bands, where we also show the trimmed Cepheids for completeness.  Figure \ref{fig:PL3} shows the final phase-averaged PL relations.  The apparent gap in the period distribution near 20 days is probably due to losing these Cepheids due to some combination of sampling and lunation.  The low scatter near 10 or 40 days shows that we are not aliasing them to half or twice that period. Comparing Figures \ref{fig:PL3rand} and \ref{fig:PL3}, we see that the dispersions about the PL relations for our phase-averaged Cepheids of 0.12, 0.08, and 0.11 mag, for the B, V and I band respectively, are about half that of the corresponding random-phase relations dispersions of 0.29, 0.17, and 0.20~mag.  The template-fit calibrated Cepheids are clearly a much better fit to the PL relations, validating our overall approach. The scatter in these PL relations are comparable to the scatter in the OGLE II Cepheids in the LMC of 0.24, 0.16, and 0.11~mag for the B, V and I band respectively.  

The variances about the PL relations are highly correlated because after fitting individual extinctions there are only 2 degrees of freedom for each Cepheid (ignoring the single, global distance variable).  We can capture this reduced error space by defining two orthogonal error vectors.  The first, $\vec{E}_1=\vec{\mu}-( \vec{\mu} \cdot \vec{R})\vec{R}/(\vec{R} \cdot \vec{R})$ corresponds to errors in distance that cannot be modeled as extinction. The second, $\vec{E}_2= \vec{\mu} \times \vec{R}$, corresponds to residuals that can be modeled neither by changes in distance nor extinction.  Here $\vec{\mu}=\{1, 1, 1\}$ corresponds to a change in distance, $\vec{R}=\{R_B, R_V, R_I\}$ corresponds to a change in extinction and $\hat{E_1}\cdot \vec{R}=\hat{E_2}\cdot \vec{R}=\hat{E_1}\cdot \hat{E_2}=0$.  When normalized, these vectors are $\hat{E}_1=-0.475\hat{b}+0.102\hat{v}+0.874\hat{i}$ and $\hat{E}_2=0.466\hat{b}-0.814\hat{v}+0.348\hat{i}$. Figure \ref{fig:e1e2} shows the residuals in terms of $\hat{E}_1$ and $\hat{E}_2$. The dispersion is dominated by $\hat{E_1}$ residuals, but the residuals are also correlated, with the $\hat{E_2}$ residuals increasing with the $\hat{E_1}$ residuals.  This shows us that the colors of the Cepheids are not completely characterized by the PL relations and extinction.  

We carried out additional tests to identify and quantify possible errors in the distance due to uncertainties in the mean magnitudes and to explore the effect of the light-curve template fitting procedures.  Since the calibration zeropoint estimates are partially degenerate with light curve amplitudes, we fit each individual light curve using the previously-determined LBT zeropoint as well as values shifted by $\pm0.3$ mag, which is about 4 times our estimated uncertainties.  The resulting mean magnitudes changed very little and the distance modulus changed by a maximum of 0.01 mag, which is smaller than our statistical uncertainties. The residuals about the PL relations were also slightly larger when we used the arbitrarily-shifted zeropoints.

We also tested how biases in the amplitude estimates would affect the distance estimate.  We refit our light curves by changing the previously-determined amplitudes by $\pm0.07$.  Recall that the mean template amplitude parameter offset we observed relative to the F94 Cepheids was +0.07 (see \S2).  The relative distance moduli determined from these altered light curve fits were $\Delta\mu_{LMC}=9.12\pm0.03$~mag for the increased amplitudes and $\Delta\mu_{LMC}=9.32\pm0.03$~mag for the decreased amplitudes.  These values correspond to offsets of $-0.07$~mag and $+0.13$~mag relative to the distance modulus from our final sample, respectively.  Since artificially increasing the amplitudes results in a smaller distance modulus, biases in amplitude estimates cannot explain the difference between our distance estimate and that of F94 (see below). While the scatter about the PL relations with the modified amplitudes did not change significantly for the B and V bands, it nearly doubled in the I band, rising from 0.11 mag to 0.20 mag for the ``increased amplitude'' fits and to 0.15 mag for the ``decreased amplitude'' fits. Equivalently, the rescaled $\chi^2$ increased by $\Delta \chi^2=381$ and $467$ for the increased and decreased amplitudes, respectively, statistically ruling out the models with shifted amplitudes.

Table~\ref{tab:distnoz} compares our estimated distance modulus to previous estimates that are also based on {\it HST} observations of Cepheids. Here we removed the different metallicity corrections and examine $\Delta\mu_{LMC}$ to remove the differing assumptions for the LMC distance modulus so that we could compare the estimates under the same assumptions.  The {\it HST} Key Project on the Distance Scale \citep[][hereafter ``KP'']{freedman2001} reported a final distance modulus based on 17 Cepheids of $\Delta\mu_{LMC}=9.25\pm0.08$~mag (subtracting their metallicity correction of 0.05 mag). This agrees with our distance of $\Delta\mu_{LMC}=9.19\pm0.02$~mag given the uncertainties. \citet[][hereafter ``ANGST'']{mccommas2009} used ACS ANGST data for a region in the outer disk of M81 where we have no LBT data.  Using a total of 20 orbits they identified 13 Cepheids (both fundamental and overtone) with periods less than 10 days and determined a distance modulus of $\Delta\mu_{LMC}=9.34\pm0.05$ mag (removing a metallicity correction of 0.03~mag).  In both cases, the primary systematic uncertainties are related to metallicity and the distance to the LMC and these uncertainties are not relevant to the present comparison. Therefore, the ANGST distance, which differs from our estimate by 0.15~mag, is not consistent given the relative uncertainties for this comparison.  Some of the difference among these estimates may arise from the use of BVI bands (in our study) instead of VI bands (KP and ANGST), the adopted PL relations (KP) or extinction (KP), issues we examine next. 

Table~\ref{tab:distnoz} also explores how our estimated distance modulus depends on the choice of bands, the number of bands, the adopted PL relations, and the period and galactic radius distributions of the Cepheids. We began by determining the distance modulus using the three possible combinations of two bands, for both the phase-averaged and random-phase magnitudes.  The distances found with phase-corrected magnitudes are systematically smaller by $\sim0.03$~mag and have smaller uncertainties than those found with random-phase magnitudes. However, the differences in the distance estimates are always mutually consistent.  There are significant differences between the permutations of the bands fit, the largest being the 0.15 mag offset between the BV and the VI fits.  If we compare our VI distance to the ANGST and KP estimates, we find offsets of $-0.12$ and $-0.03$~mag, respectively, in the sense that our distance is smaller.  We are consistent with the KP distance given the uncertainties, but we still differ by $2\sigma$ from the ANGST distance for the same assumed LMC distance and assuming no metallicity correction. The KP used slightly different PL relations from the original \citet{udalski1999} paper rather than the updated PL relations associated with the catalog release. If we fit our VI data with the original OGLE II PL relations, the distance modulus increases from $\Delta\mu_{LMC}=9.22\pm0.03$ to $9.25\pm0.02$, which is equal to the KP estimate rather than 0.03 mag smaller.  \citet{mccommas2009} used short-period Cepheids, $P<10$ days, and included both fundamental and overtone pulsators.  While we have no overlap with their period range, we can divide our sample in period to see if there are any effects.  We divided the sample at the median period of 21 days and determined the distance moduli for each subset. The two subsamples give distance moduli differing by $\Delta\mu=0.02\pm0.03$~mag, with the long-period Cepheids yielding a slightly larger distance modulus.  In Figure \ref{fig:wmag}, we reproduce Figure 8 from \citet{mccommas2009}, adding our Wesenheit magnitudes and the $W_{VI}$ PL relation from Equation~\ref{eqn:pls}.  Our Wesenheit distance modulus of $\Delta\mu_{LMC}=9.22\pm0.03$ mag falls between those of the KP ($\Delta\mu_{LMC}=9.14\pm0.07$ mag) and \citet{mccommas2009} ($\Delta\mu_{LMC}=9.37\pm0.05$ mag).  %WESENHEIT paragraph
 These two band fits to determine distances and extinction of individual Cepheids are nearly equivalent to using the Wesenheit magnitudes which implicitly correct for individual extinctions.  The only significant difference in our approach is that the fit for each band is weighted by the photometric error in that band. If we weight the bands equally, then the results are identical to using Wesenheit magnitudes.  In doing these fits, we set the photometric error so that the error in the individual distances would be the same as when using the true photometric uncertainties.  When we redo the two band fits using equal photometric errors for the two bands, so as to mimic the Wesenheit approach, the distance moduli change by less than 1$\sigma$.  Therefore, we do not report all the individual results.

The systematic offsets between the filter combinations and the lack of difference with period are suggestive of an inconsistency in the color of the PL relations in Eqn.~\ref{eqn:pls}. We refit the data allowing an adjustment to the B- and V-band PL zeropoints while holding the I band fixed. We set a prior to keep the shifts small and found B- and V-band corrections of $+0.02\pm0.02$ mag and $-0.01\pm0.01$, respectively.  The bootstrap resampling results yielded a reduced uncertainty on the B-band correction of $+0.02\pm0.01$ mag and no changes in the V-band correction relative to the $\chi^2$ estimate.  These corrections are inconsistent with zero because of the structure of the error ellipse --- the value of zero for the two parameters is ruled out at $3.5\sigma$. These small PL zeropoint shifts bring the VI, BV and BI distance moduli into agreement at $\Delta\mu_{LMC}=9.21\pm0.03$, $9.19\pm0.03$ and $9.19\pm0.02$~mag.  Using bootstrap re-sampling, we find $\Delta \mu_{LMC} = 9.21\pm0.04$, $9.19\pm0.05$, $9.19\pm0.03$~mag, respectively. The shifts could be due to small offsets in the absolute calibration of the (ground-based) OGLE and (space-based) ANGST photometry.  Another possibility is that the difference stems from a metallicity effect and can be described by taking the galactocentric position of the Cepheids into account. For example, while we are generally consistent with the KP results, whose fields overlap ours, the \citet{mccommas2009} field is so far from the center of M81 as to be outside our field of view.  We explore this in the next section.

\section{Evidence for A Systematic Dependence on Radius}

We next looked for any correlations of the PL residuals or extinction with galactocentric position as a proxy for metallicity.  We estimated the deprojected radius using an inclination angle of $i=59\degr$ and a major-axis position angle of PA=$157\degr$~(\citealt{kong2000}).  The center of M81 lies at R.A.=09h55m33.1730s, Dec.=+69d03m55.061s (J2000.0) based on \citet{johnston1995}.  If we define the $x$ axis to lie along the major axis, and the $y$ axis to lie along the minor axis, the deprojected radius is simply $\rho=(x^2+y^2/\cos^2i)^{1/2}$.

\citet{zaritsky1994} derived an abundance gradient for M81 of $[O/H]=12+\log(O/H)=-0.12\pm0.05~{\rm dex}/\rho_{s}$, where $[O/H]=9.10\pm0.11$~dex at $0.8\rho_{s}$ and $\rho_s$=$2\farcm94$ is the scale radius of the galaxy. They used abundances for 26 \ion{H}{2} regions over the radius range $1.2\rho_s < R < 3.6\rho_s$. Our Cepheids have galactocentric radii from $1.1\rho_{s}$ to $4.1\rho_{s}$, corresponding to a range in metallicity of $[O/H]=9.06$ to $8.70$~dex. This is a wide range of abundances for a single Cepheid sample. Given a ``typical'' metallicity correction of $\gamma_{VI}=-0.24$~mag dex$^{-1}$ \citep{kenni1998,sakai2004} relative to an abundance of $[O/H]=8.5$~dex, we would expect to see a significant trend in our sample as a function of galactocentric radius, ranging from $+0.13$~mag at $[O/H]=9.06$~dex to $+0.05$~mag at $[O/H]=8.70$~dex. We also note that the projected radius of the \citet{mccommas2009} field lies at $\rho \sim 5\rho_s$, corresponding to [O/H]$\simeq$8.6 and a metallicity correction of $+0.02$~mag.  The KP fields have a typical radius of $\sim 2.1\rho_s$ corresponding to $[O/H]\sim 8.94$~dex and a metallicity correction of $+0.11$~mag.  This is similar to our mean radius of $2.7\rho_s$.

Figures \ref{fig:resid},  \ref{fig:bvir}, and \ref{fig:exr} show three views of the radial dependence of the residuals.  Figure \ref{fig:resid} shows the distribution of the PL residuals with radius.  The trends depend on the band: the B band shows a slightly negative slope, while the V and I bands both have a slightly positive slope.   This suggests that the Cepheids are becoming bluer with increasing radius.  In Figure \ref{fig:bvir}, we examine the extinction corrected colors, and we indeed see the Cepheids become bluer with increasing radius.  This is the same sense expected if line blanketing makes the more metal-rich inner stars redder or if metal-rich Cepheids are simply brighter.  In this latter case, small changes in extinction are used to compensate for the increased luminosity, making the inner Cepheids appear redder.  Finally, in Figure \ref{fig:exr} we examine the distribution of extinctions and the $\hat{E}_1$ and $\hat{E}_2$ residuals.  All three of these parameters increase with radius.  The strongest trend is in extinction, although the trend is in the opposite direction expected if metallicity were causing the inner Cepheids to be redder or brighter than the outer Cepheids. Trends here are hard to interpret since the mean extinction is also likely a function of galactocentric radius.  The trend in $\hat{E_2}$ suggests a color dependence on radial position, although the stronger trend in $\hat{E_1}$ suggests that the effects are dominated by luminosity rather than color. 

We investigated this further by dividing the data into 3 radial bins, each containing one-third of the Cepheids. As shown in Table~\ref{tab:distnoz}, the estimated distance moduli are statistically consistent.  There are, however, differences between the middle Cepheids when compared to the inner and outer bins. While the inner two subsamples follow the expected trend of increasingly metal-rich Cepheids being brighter and leading to an underestimate of the distance, the trend is not smooth.  We also fit the 3 permutations of two-band distance moduli estimates for each of these radial bins.  The differences are largest for the inner two bins, while the distance moduli for the outermost bin agree within $1.4\sigma$. For the metallicities and gradients reported by \citet{zaritsky1994}, the outer bin has a metallicity of $[O/H]=8.8$~dex, comparable to that of the LMC ($[O/H]=8.5$~dex) where the fiducial PL relations were determined \citep{udalski1999}. This suggests that the differences seen between the $VI$-band fits in \S\ref{sec:pls} are unlikely to be due to any photometric offsets between the OGLE and ANGST data sets, but are instead caused by the different abundances of the Cepheid samples.

Next we add a radius dependence to the zeropoints in Equation~\ref{eqn:pls} of the form 
\begin{equation}
\label{eqn:rad}
-(\gamma_\mu\hat{\mu} + \gamma_2\hat{E_2}) \left( \frac{\rho-6.0\rho_s}{\rho_s/0.12}\right).
\end{equation}
We include no metallicity term proportional to $\vec{R}$ because it cannot be measured without independent constraints on the extinction.  We chose the outer radius of $6\rho_s$ to roughly correspond to the radius where the metallicity will match that of the LMC.  While these are strictly fits in radius, the slope is scaled by $-\rho_s/0.12$ so the values we find for $\gamma_\mu$ and $\gamma_2$ are the metallicity dependence in mag dex$^{-1}$ given the \citet{zaritsky1994} estimate of the metallicity gradient.  With these additional terms we find a distance modulus of $\Delta\mu_{LMC}=9.39\pm0.08$~mag that is 0.19~mag ($2.3\sigma$) larger than the estimate without the correction.  The fit parameters are $\gamma_\mu=-0.56\pm0.21$~mag dex$^{-1}$ and $\gamma_2=0.07\pm0.03$~mag dex$^{-1}$. Using the bootstrap method we find $\Delta\mu_{LMC}= 9.39\pm0.14$~mag, $\gamma_\mu=-0.56\pm0.36$~mag dex$^{-1}$ and $\gamma_2=0.07\pm0.03$ mag dex$^{-1}$.  For comparison, the bootstrap distance modulus without metallicity terms is $\Delta \mu_{LMC} = 9.19\pm0.03$~mag.  Figure \ref{fig:boot3} shows in blue the likelihood contours enclosing 68.3\% and $95.4$\% of the bootstrap trials.  The small but statistically significant value of $\gamma_2$ indicates that the dominant effect is a zeropoint shift (i.e. a change in luminosity), but that the color effect is non-zero.  Remember, however, that there can be a second color term degenerate with extinction that we cannot measure.  As shown in Table~\ref{tab:distz}, when we apply these corrections to the three possible 2-band fits, the estimated distance moduli in the inner two bins come into much better agreement, although the distance moduli of the outer bin are slightly less consistent.  Nonetheless, the distance moduli in each bin agree within $1.6\sigma$ based on the $\chi^2$ results, and at the $1\sigma$ level with the uncertainties estimated from bootstrap resampling as compared to the  $5.1\sigma$ ($\chi^2$) and $3.1\sigma$ (bootstrap) disagreement without these terms.  We adopt the bootstrap results, including the uncertainties of the metallicity dependence, for this estimate of the distance of M81. This metallicity-corrected distance modulus of $\Delta\mu_{LMC}= 9.39\pm0.14$~mag agrees with the metallicity-corrected results of both \citet{freedman2001} and \citet{mccommas2009} but is more statistically robust.

\section{Metallicity and Calibration to NGC~4258}

If NGC~4258 is to be the distance calibrator instead of the LMC, we must use a common metallicity correction to the LMC PL relations in order to self-consistently determine the distance moduli and the metallicity corrections. When we combine datasets from two galaxies, we again note that it is the relative abundance and not the absolute abundance that is important.  We use the metallicity gradient for NGC~4258 from \citet{zaritsky1994}, as did \citet{macri2006}, so that the metallicity gradients for both galaxies are from the same source and on the same scale. As we will later emphasize, it is also only possible to do this correctly as a joint fit to the three systems. 

We used the positions and mean B, V, and I magnitudes of the 69 Cepheids in the final sample of \citet{macri2006}.  We redetermined the deprojected radii using the same method we used with M81 with an inclination angle of $i=150\degr$ and a major-axis position angle of PA=$72\degr$~\citep{vanAl1980}.  We first determined the distance modulus and metallicity correction from the NGC~4258 data alone.  Matching the structure of Equation~\ref{eqn:rad} for M81, the metallicity correction for NGC~4258 is, 
\begin{equation}
-(\gamma_\mu\hat{\mu} + \gamma_2\hat{E_2}) \left( \frac{\rho-4.85\rho_0}{\rho_0/0.14}\right),
\end{equation}
where we again centered the metallicity correction at the radius where the metallicity will match that of the LMC, $\rho_0$=$2\farcm22$, and scaled the slope so the values we find for $\gamma_\mu$ and $\gamma_2$ are the metallicity dependence in mag dex$^{-1}$ given the estimate of the metallicity gradient.  Fitting only NGC~4258, we found $\mu_{N4258}-\mu_{LMC}=10.89\pm 0.04$~mag, with a metallicity correction of $\gamma_\mu=-0.33\pm0.09$~mag dex$^{-1}$ and $\gamma_2=0.01\pm0.02$~mag dex$^{-1}$ and bootstrap resampling results of $\mu_{N4258}-\mu_{LMC}=10.89\pm 0.04$~mag,  $\gamma_\mu=-0.33\pm0.10$~mag dex$^{-1}$ and $\gamma_2=0.01\pm0.02$~mag dex$^{-1}$.  These results, shown in green in the left panel of \ref{fig:boot3}, agree with those from \citet{macri2006} of $\mu_{N4258}-\mu_{LMC}=10.88\pm 0.04$~mag, and $\gamma_\mu=-0.29\pm0.09$~mag dex$^{-1}$. \citet{macri2006} did not allow any color terms to the metallicity dependence and so they forced $\gamma_2\equiv0$. 

Next we simultaneously fit the M81 and the NGC~4258 samples to find that $\mu_{M81}-\mu_{LMC}=9.33\pm 0.03$~mag, $\mu_{N4258}-\mu_{LMC}=10.90\pm 0.03$~mag and $\mu_{N4258}-\mu_{M81}=1.57\pm0.04$~mag, with metallicity corrections of $\gamma_\mu=-0.36\pm0.08$~mag dex$^{-1}$ and $\gamma_2=0.02\pm0.01$~mag dex$^{-1}$.  The bootstrap resampling results are $\mu_{M81}-\mu_{LMC}=9.33\pm 0.05$~mag, $\mu_{N4258}-\mu_{LMC}=10.90\pm 0.04$~mag, $\mu_{N4258}-\mu_{M81}=1.57\pm0.04$~mag, $\gamma_\mu=-0.36\pm0.10$~mag dex$^{-1}$ and $\gamma_2=0.02\pm0.01$~mag dex$^{-1}$.  We adopt the more conservative results from bootstrap resampling.  The joint results largely agree with those from the analyses of the separate galaxies.  The left panel in Figure \ref{fig:boot3} shows contours from the bootstrap resampling results for the estimates of the metallicity parameters.  The metallicity corrections from the combined dataset fall between the estimates for the two galaxies, but are dominated by NGC~4258 due to the larger metallicity range ($\Delta[O/H]=0.54$~dex versus $\Delta[O/H]=0.36$~dex) implied by the metallicity gradients. The left panel in Figure \ref{fig:dist} similarly shows contours for the distance estimates.

In fact, the uncertainties in the relative metallicities and their gradients are crucial to the analysis, but have not been formally included in Cepheid studies.  \citet{bono2008}, \citet{bresolin2011a} and \citet{shappee2010} have recently pointed out that they are important, and rescale their distances and metallicity corrections for the changing estimates of the gradients.  But no one has simply included these uncertainties with those of the Cepheids.  One problem is that published fits for the metallicities and their gradients do not include the full covariance matrix of uncertainties needed to correctly include them in an analysis.   Here we solve this problem by bootstrap resampling the metallicity gradient.  
We extracted the $R_{23}$ [O/H] abundance estimates for M81 and NGC~4258 from \citet{zaritsky1994}.  If we estimate the metallicity gradients using bootstrap resampling to determine the errors, we find $[O/H]=(8.92\pm0.01)-(0.13\pm0.03)(\rho-2.15\rho_s)$ for M81 and $(8.90\pm0.03)-(0.08_{-0.04}^{+0.03})(\rho-2.09\rho_0)$ for NGC~4258, which are consistent with \citet{zaritsky1994}.  Here, we have centered the relations at the average radius of the \ion{H}{2} regions in order to minimize the error covariances.  We then redetermined the distance and metallicity parameters by bootstrap resampling over both the \ion{H}{2} regions and the Cepheids and further included the uncertainties in the LMC metallicity of $8.5\pm0.08$ dex \citep{ferrarese2000}.  Operationally, we randomly set the LMC metallicity as 8.5 plus a $\sigma=0.08$ dex Gaussian random number, randomly resampled the \ion{H}{2} regions of the galaxies, fit the metallicity gradients, randomly resampled the Cepheids and finally refit for the Cepheid distances and metallicity parameters given the estimated metallicity gradient and the random sample of the Cepheids.

If we fit the individual galaxies, we find a metallicity correction of $\gamma_\mu=-0.45_{-0.41}^{+0.38}$~mag dex$^{-1}$ and $\gamma_2=0.08_{-0.04}^{+0.05}$~mag dex$^{-1}$ with a distance modulus of $\mu_{M81}-\mu_{LMC}=9.32_{-0.12}^{+0.15}$~mag for M81, and a metallicity correction of $\gamma_\mu=-1.88_{-1.44}^{+0.80}$~mag dex$^{-1}$ and $\gamma_2=0.001\pm0.013$~mag dex$^{-1}$ and a distance modulus of $\mu_{N4258}-\mu_{LMC}=11.75_{-0.40}^{+0.70}$~mag for NGC~4258.  If we compare these results to those obtained without considering the uncertainties of the metallicity gradients, we find significant changes in the metallicity correction parameter $\gamma_{\mu}$, 0.11~mag dex$^{-1}$ smaller for M81 and 1.55 mag dex$^{-1}$ larger for NGC~4258.  The distance moduli for M81 and NGC~4258 changed by $-0.07$~mag and $+0.86$ mag, respectively.  The problem for NGC~4258 is that the metallicity gradient is very sensitive to exactly which \ion{H}{2} regions are included, as noted by \citet{bono2008}.  Most importantly, the range of possible gradients from the bootstrap analysis includes solutions near zero slope.  With a metallicity gradient near zero, the metallicity parameters, $\gamma_\mu$ and $\gamma_2$, will diverge and the distance cannot be well determined.  Thus, we see in Figure \ref{fig:boot3} and Figure \ref{fig:dist} that when we allow for uncertainties in the metallicity gradients, M81 is a better calibrator for metallicity effects than NGC~4258. Figure \ref{fig:dist} also shows us that the uncertainty in distance for $\mu_{N4258}-\mu_{M81}$ is smaller than either distance relative to the LMC because the metallicities of NGC~4258 and M81 overlap, reducing the effects of the metallicity corrections. 

%new bresolin paragraph
 The metallicity gradient of NGC~4258 was revised by \citet{bresolin2011a} with 36 new measurements of \ion{H}{2} regions and other measurements from the literature. The resulting metallicity gradient is shallower at $-0.051$~dex$/\rho_0$ and results in a metallicity correction of $\gamma_\mu=-0.69$~mag dex$^{-1}$ when applied to the \citet{macri2006} Cepheids \citep{bresolin2011a}. 
  In order to include this new data in our determination of the distance and metallicity correction, the metallicities of the \ion{H}{2} regions must all be determined using the same calibration. Unfortunately, \citet{bresolin2011a} does not supply the necessary information to transform between the \citet{zaritsky1994} and \citet{mcgaugh1991} calibrations self-consistently.  We attempted to do so by shifting the \citet{bresolin2011a} data calibrated with \citet{mcgaugh1991} by the mean offset from the \citet{zaritsky1994} data.  This puts the data on the $R_{23}$ scale of \citet{zaritsky1994}.  We discuss how to transform from this metallicity scale to the $T_e$ scale of \citet{bresolin2011a} in \S7.
With the \citet{bresolin2011a} data added, the potentially problematic \ion{H}{2} region discussed above has a much smaller impact, and we find a metallicity correction of $\gamma_\mu=-0.77_{-0.32}^{+0.27}$~mag dex$^{-1}$ and $\gamma_2=0.00\pm0.01$~mag dex$^{-1}$.  This is smaller and much better constrained than that found with the \citet{zaritsky1994} data alone.  As expected, if we instead shift the \citet{zaritsky1994} data to the \citet{mcgaugh1991} scale, we find the same metallicity correction, but a slightly different distance modulus due to the shifted mean metallicity. Figures \ref{fig:boot3} and \ref{fig:dist} show the results of including the \ion{H}{2} regions from both \citet{bresolin2011a} and \citet{zaritsky1994}, after shifting them to the \citet{zaritsky1994} scale we must use for M81.

  When we combine the M81 and NGC~4258 datasets, we find distances of $\mu_{M81}-\mu_{LMC}=9.40_{-0.11}^{+0.15}$~mag, $\mu_{N4258}-\mu_{LMC}=11.08_{-0.17}^{+0.21}$~mag and $\mu_{N4258}-\mu_{M81}=1.68\pm0.08$~mag, with metallicity corrections of $\gamma_\mu=-0.62_{-0.35}^{+0.31}$~mag dex$^{-1}$ and $\gamma_2=0.01\pm0.01$~mag dex$^{-1}$.  The left panel in Figure \ref{fig:boot3} shows contours of the metallicity parameter estimates from these bootstrap resampling results.  The solid lines use only the \citet{zaritsky1994} data while the dotted lines used the combined \citet{zaritsky1994} and \citet{bresolin2011a} datasets.  We can see a change from our previous results in the right panel, although the parameters from the joint M81, NGC~4258 dataset including the metallicity gradient uncertainties are consistent with the previous results. Note the strong covariance of the distances in Figure \ref{fig:dist}.  The distance from M81 to NGC~4258 is relatively tightly constrained because the metallicities overlap independent of their uncertainties.  The distances between the two galaxies and the LMC are far more uncertain because they depend strongly on the uncertain metallicity gradients.  We find very little change when we include the \citet{bresolin2011a} data shifted to the \citet{zaritsky1994} gradient intercept, as the metallicity gradient of M81 still dominates.

Figure \ref{fig:dist} shows the uncertainty in the ``geometric'' estimates of the distances to the three galaxies.  We use the \citet{bartel2007} M81 distance modulus of $27.99\pm0.16$~mag determined using the Expanding Shock Method (ESM) for SN1993J, the NGC~4258 maser distance modulus of $29.29\pm0.15$~mag \citep{herrn1999} and an LMC distance modulus of $18.5\pm0.1$~mag \citep{freedman2001} that is consistent with recent estimates from eclipsing binaries (see \citealt{bonanos2011}).  
Adding these constraints as a prior has little effect on the results because the statistical weights of the Cepheid and metallicity data are so much greater and because our approach does not allow the distance prior to constrain the metallicity gradients.  If we did so, we would find that the flat metallicity gradients driving some of these uncertainties would be ruled out. 
Our Cepheid results are generally consistent with these independent distance estimates, but suggest that the ESM distance to M81 may be somewhat high. 

Figure \ref{fig:met} compares our estimate of the metallicity dependence to previous work. \citet{kochanek1997} used 17 galaxies in multiple bands and found a metallicity dependence of $\gamma_\mu=-0.14\pm0.14$~mag dex$^{-1}$ along with a correction for a color dependence, $\gamma_V-\gamma_I=0.13\pm0.04$~mag dex$^{-1}$.  \citet{kenni1998} used HST to study 2 fields within M101, finding $\gamma_\mu=-0.24\pm0.16$~mag dex$^{-1}$.  \citet{sakai2004} also found $\gamma_\mu=-0.24\pm0.05$~mag dex$^{-1}$ by comparing Cepheid distances to tip of the red giant branch (TRGB) distances for 17 galaxies. \citet{groen2004} used Galactic Cepheids with individually determined metallicities and distances and found $\gamma_\mu=-0.6\pm0.4$~mag dex$^{-1}$.  When 5 Magellanic Cloud Cepheids were added to the sample, they found $\gamma_\mu=-0.27\pm0.08$~mag dex$^{-1}$.  Determinations based on the comparison of two fields within a single galaxy were carried out by \citet{macri2006}, \citet{scowcroft2009} and \citet{shappee2010} found $\gamma_\mu=-0.29\pm0.09,-0.29\pm0.11$ and $-0.83\pm0.21$~mag dex$^{-1}$ for NGC~4258, M33 and M101, respectively. Many of these differences are either explicitly or implicitly due to differences or uncertainties in the metallicities and their gradients, rather than an issue fundamental to Cepheids.  We also display our results for the fits using the combined \citet{zaritsky1994} and shifted \citet{bresolin2011a} data for the NGC~4258 \ion{H}{2} regions labeled with ``Revised''.

\section{Discussion}

The importance of Cepheids for determining distances and the Hubble constant makes it crucial that we understand the systematic uncertainties associated with this standard candle. The principle concerns are the absolute calibration of the distance scale, and the effects of metallicity and blending.  To properly investigate the systematic problems, we need large samples of Cepheids in environments of varying abundance with data in bands that both maximize and minimize the effects of metallicity. Having data in more than two photometric bands is the minimum necessary to begin separating the effects of extinction and metallicity.  In this study, we have used LBT V-band observations over a long temporal baseline to identify and phase Cepheids in M81 and (generally) single-epoch HST photometry from the ANGST project (\citealt{dalcanton2009}) to provide HST-calibrated mean BVI magnitudes. 

The combination of ground-based monitoring with space-based calibration data optimizes the use of telescope time and maximizes the number of Cepheids. This project used 72 archival orbits of HST time and produced 107 calibrated Cepheid light curves in 3 bands, while \citet{freedman1994} used 48 orbits of HST time to identify 30 Cepheids, of which only 17 were used for the final distance determination \citep{freedman2001}. Crudely, our approach uses HST 6 times more efficiently than the traditional approach, and we expect that to rise as we use subsequent LBT epochs to increase the yield of Cepheids. The one weakness in our approach is that determining the light curve amplitudes from single calibrating epochs is clearly risky, and it would be best to have at least two HST epochs in one band to better constrain and check the amplitudes.  Since amplitudes at other bands are determined by the calibrating band \citep[e.g.][]{stetson1996,yoachim2009,tanvir2005,freedman2010}, this is not necessary for any additional bands. In addition to our more efficient use of HST time, we also have more accurate periods due to our longer baseline and larger number of epochs.  For example, we note that the \citet{freedman1994} sample has a fractional period error of $\langle \Delta P/P \rangle=+0.03$, where their periods were systematically shorter.  This ``period bias'' leads to a distance error of $\Delta \mu \backsimeq -1.5\langle \Delta P/P \rangle$ $\sim-0.05$~mag for the V and I band observations.  If such errors are typical or, worse yet, systematic for sparsely-sampled Cepheid light curves, this represents a significant problem for the traditional approach if the goal is a ~1\% local calibration of the distance scale. 

We fit our final sample of 107 Cepheids to the OGLE II BVI PL relations and determined a relative distance modulus to M81 of $\Delta \mu_{LMC} = 9.19\pm0.02$~mag ($\chi^2$) and $\Delta \mu_{LMC} = 9.19\pm0.03$~mag (bootstrap) if we ignore any metallicity dependance.  This result is consistent with \citet{freedman2001}, but significantly smaller than that of \citet{mccommas2009} and already has residuals comparable to the OGLE II PL relations for the LMC.  We also fit the data for all four permutations of the three bands, essentially corresponding to using Wesenheit magnitudes, and found that the results were not mutually consistent.  This suggests that the colors of the PL relations need to be corrected.  Fixing the I band PL, we found that PL relation zeropoint corrections to the B and V bands of $+0.02\pm0.02$~mag and $-0.01\pm0.01$~mag, respectively, would reduce these discrepancies. Due to the shape of the error ellipse, a correction of zero is ruled out at 3.5$\sigma$. Dividing the sample into bins by period and radius showed that the inconsistencies were a function of radius. In particular, the two-band fits of the inner radial bins, where the metallicities have the largest differences from the calibrating sample in the LMC, were the most mutually inconsistent.  We solved for a radius-dependent zeropoint, $\gamma_\mu$, and a color term, $\gamma_2$, orthogonal to extinction and distance.  We found corrections of  $\gamma_\mu=-0.56\pm0.36$~mag dex$^{-1}$ and $\gamma_2=0.07\pm0.03$~mag dex$^{-1}$, assuming the \citet{zaritsky1994} estimate of the abundance gradient.  When we apply these position-dependent zeropoint shifts, we find a metallicity-corrected distance modulus of $\Delta\mu_{LMC}= 9.39\pm0.14$~mag which agrees with both the \citet{freedman2001} and \citet{mccommas2009} results.  We also find that the two-band fits made with these corrections are internally consistent within each radial bin. We cannot easily compare our estimate of $\gamma_2$ to the only earlier estimate of a color correction by \citet{kochanek1997}, although our correction appears to be smaller.  Our metallicity correction $\gamma_\mu$ is somewhat larger than past estimates, but is in general agreement given the uncertainties.  

We next tried to self-consistently solve for the relative distances of the LMC, M81 and NGC~4258, including the metallicity uncertainties.  If we assume, as is often done, that the metallicity gradients have no uncertainty, then NGC~4258 dominates the estimate of the metallicity correction and we find $\gamma_\mu=-0.36\pm0.10$~mag dex$^{-1}$ and $\gamma_2=0.02\pm0.01$~mag dex$^{-1}$.  However, given the increasing evidence that the uncertainties in the relative metallicities are crucial to the estimate of the metallicity corrections to the Cepheid distance scale \citep{bono2008,shappee2010, bresolin2011a}, we redid the models including a fit to the metallicities of the \ion{H}{2} regions as part of the analysis.  When this is included, NGC~4258 is a far poorer calibrator for metallicity effects than M81 because it is possible for its metallicity gradient to be very shallow.  The revised metallicity gradient of \citet{bresolin2011a} improves this.  However to accurately know the full effect of the new gradient, we need the metallicities of all the \ion{H}{2} regions, found with a common calibration, in all the target galaxies, including the LMC.  Galaxy by galaxy revisions are of little help without a global correction procedure between different systems.  In the joint fits including the metallicity uncertainties, we find  $\gamma_\mu=-0.62_{-0.35}^{+0.31}$~mag dex$^{-1}$ and $\gamma_2=0.01\pm0.01$~mag dex$^{-1}$.  These are on the \citet{zaritsky1994} $R_{23}$ metallicity scale, which is related to the $T_e$ metallicity scale of \citet{bresolin2011a} by $Z(T_e)= a*Z(R_{23})+b$, where $a=0.69\pm0.06$ and $b=2.30\pm0.51$. The Cepheid metallicity corrections on the two scales are related by $\gamma (T_e)=\gamma(R_{23})/a=\gamma(R_{23})/(0.69\pm0.06)$ because the data actually constrain the variable combination $\gamma(R_{23})(Z(R_{23})-Z_{LMC}(R_{23}))=\gamma(T_{e})(Z(T_{e})-Z_{LMC}(T_{e}))$.

These experiments lead to two important conclusions.  First, as also recently discussed by \citet{bono2008}, \citet{shappee2010} and  \citet{bresolin2011a}, metallicities and their gradients are as important a source of uncertainties as the Cepheids.  Not in the sense of absolute abundances, but in ensuring that they are all on the same relative scale and that their uncertainties are properly included in the analysis and that the metallicities of all the \ion{H}{2} regions used are reported.  Second, as emphasized by \citet{gould1994}, \citet{hut1995}, \citet{kochanek1997} and \citet{riess2009}, the best way to analyze the Cepheid data is to simultaneously fit all the data rather than trying to reduce the problem of Cepheid distances to individual galaxies.  The latter procedure will both conceal strong, hidden distance covariances and weaken the ability to constrain sources of systematic error.  

Table~\ref{tab:distcomp} compares our results for the distance to M81 found using only the M81 data, as well as the global solution both with and without the uncertainties in the abundance gradients, to several estimates of the distance modulus of M81 based on other methods.  The ``updated uncertain gradient'' is the fit including the \ion{H}{2} regions from both \citet{bresolin2011a} and \citet{zaritsky1994}, after shifting them to the \citet{zaritsky1994} scale we must use for M81.  The absolute normalization of the distance scale is not important for this comparison, so we corrected the previous results to a common LMC distance modulus of $\mu_{LMC}=18.41\pm0.10_r\pm0.13_s$~mag, found by \citet{macri2006} using the maser galaxy NGC~4258 as the anchor of the Cepheid distance scale.  Using only M81 data, we obtain a final metallicity-corrected Cepheid distance modulus of $\mu_{M81}=27.80\pm0.14$~mag, which corresponds to a distance of $D=3.6\pm0.2$~Mpc. The global solution which accounted for the uncertainties in the abundance gradients gives an M81 distance modulus of $\mu_{M81}=27.81_{-0.11}^{+0.15}$~mag.  These results are consistent at the $1\sigma$ level with the TRGB distance of \citet{sakai2004} and the surface brightness fluctuations (SBF) distance of \citet{jensen2003}. We also agree within the errors with \citet{bartel2007}, who used SN1993J to determine a geometric distance based on ESM. The global solution which did not account for the uncertainties in the abundance gradients gave an M81 distance modulus of $\mu_{M81}=27.74\pm0.05$~mag, which is consistent with the SBF and ESM estimates, but $2.2\sigma$ from the TRGB distance estimate. 

We are expanding on this work in two dimensions.  First, we have carried out HST observations of M81 to obtain H-band photometry of most of these Cepheids.  The added bandpass will further help to separate and measure the effects of chemical composition, blending, and extinction.  Parallel observations will add a second epoch of V-band data for most of the galaxy, which will mitigate any concerns about estimating amplitudes.  The data can be further combined with the Spitzer mid-IR observations of M81 Cepheids \citep[Spitzer Proposal \#60010 by][]{freedman2008}.  Second, we are continuing to add epochs of LBT data for M81 and the other 24 galaxies that are part of the LBT monitoring program, including the maser galaxy NGC 4258. These data will enable us to identify many more Cepheids in a still broader range of environments, better characterize the systematics of the Cepheid Distance Scale, and more accurately determine the distances to these nearby galaxies.  Given the number of Cepheids we are able to identify, we are limited by systematics and not by sample size.  

\acknowledgments 

Acknowledgments: 
We would like to thank the ANGST collaboration, especially Julianne Dalcanton and Benjamin Williams, for providing B,V and I band {\it HST} photometry and useful suggestions. We would also like to thank Benjamin Shappee and Dorota Szczygie\l~for helpful discussions. 

This work is partly based on data obtained during LBC Science Demonstration Time, and we thank the instrument team and the scientists who led that effort.

This research was supported by NSF grant AST-0908816 and by HST grant AR-11760. 

The Large Binocular Telescope is an international collaboration among institutions in the United States, Italy and Germany.  The LBT Corporation partners are: the University of Arizona on behalf of the Arizona university system; the Instituto Nazionale di Astrofisica, Italy; the LBT Beteiligungsgesellschaft, Germany, representing the Max Planck Society; the Astrophysical Institute Potsdam, and Heidelberg University; the Ohio State University; and the Research Corporation, on behalf of the University of Minnesota, University of Notre Dame and University of Virginia.

This work is based on observations with the Advanced Camera for Surveys on board the NASA/ESA {\it Hubble Space Telescope}, obtained from the data archive at STScI, which is operated by AURA, Inc., under NASA contract NAS 5-26555.  

{\it Facilities:}  \facility{LBT} \facility{HST}
}

\begin{figure}
\plotone{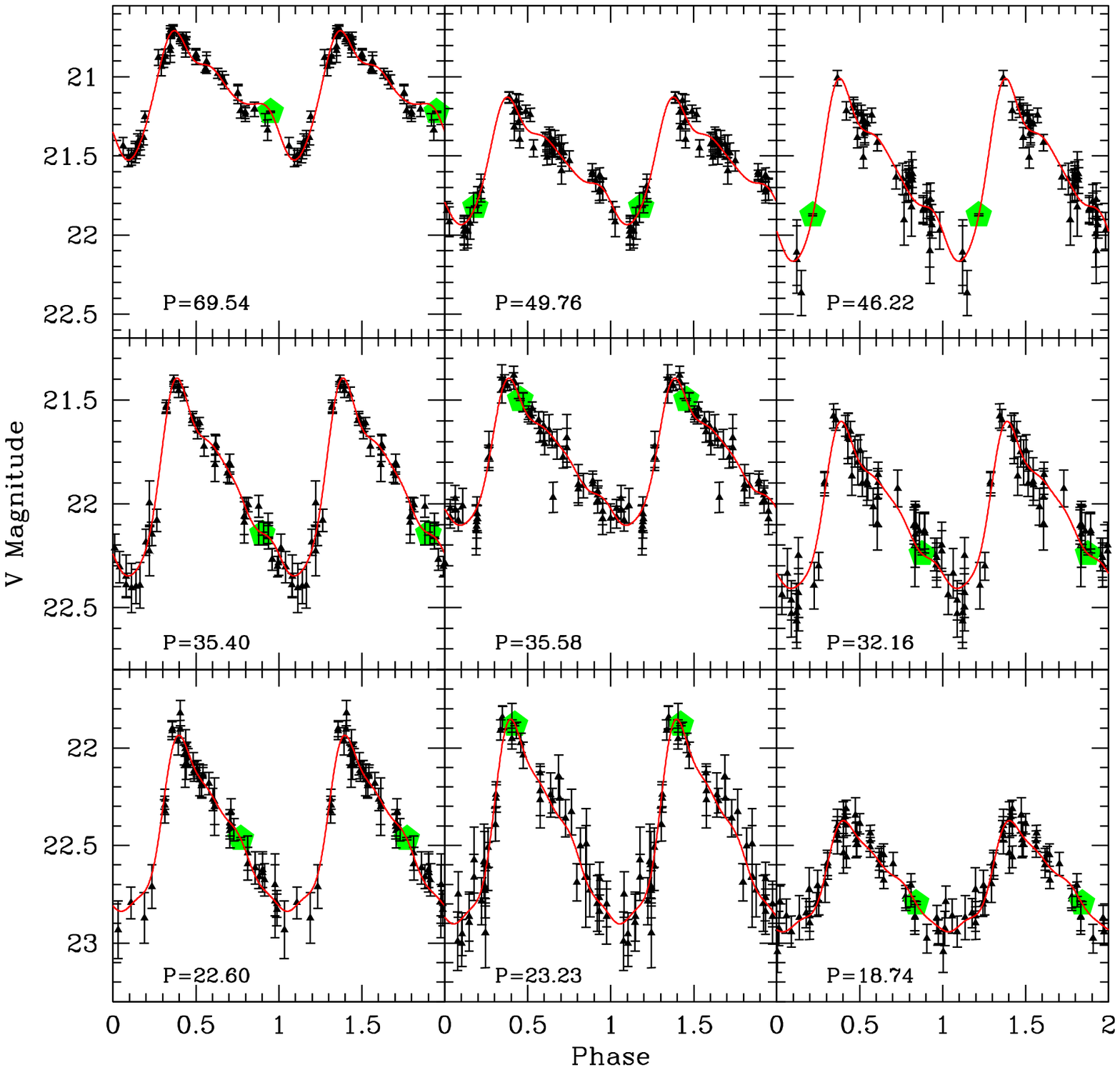}
\figurenum{1}
\label{fig:lcex}
\caption{Examples of extinction-corrected LBT V-band light curves.  The black triangles are the data and the red line is the template model fit. Within each row, the light curves range from higher to lower quality from left to right. Three separate period ranges (from longer to shorter periods) are plotted from top to bottom. The period of each Cepheid (in days) is stated in each panel. The HST calibration point is shown as a large green pentagon.}
\end{figure}

\begin{figure}
\plotone{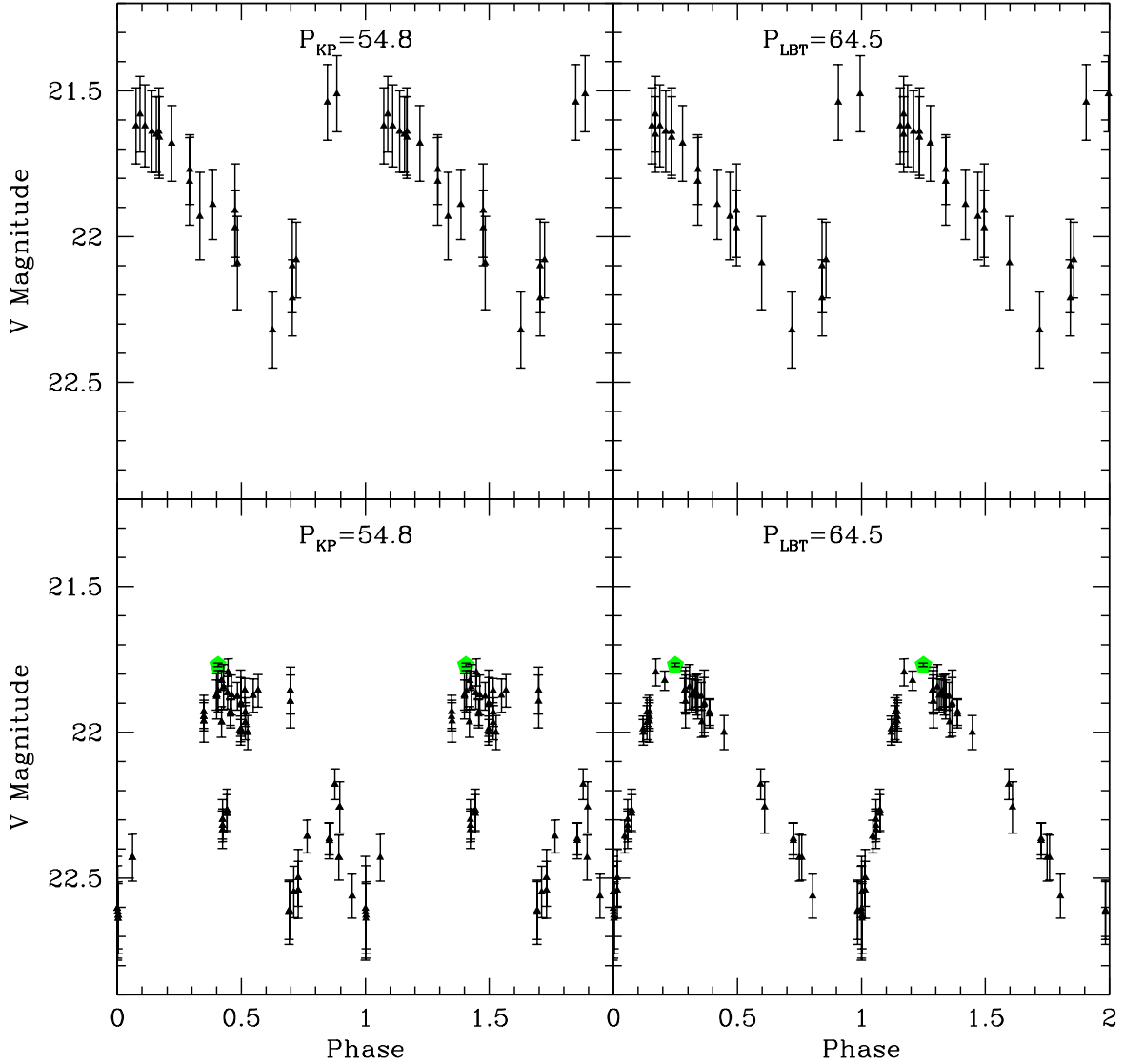}
\figurenum{2}
\label{fig:KPceph}
\caption{Phased light curves for F94-C26/M81C~095610.62+690732.7.  The top (bottom) panels show the F94 (LBT) data.  The left panels are phased to the F94 period of P=54.8 days, while the right panels are phased to the P=64.5 days period determined with the LBT data. While either period works for the F94 data, the longer LBT period is clearly correct.  The HST calibration point for the LBT light curve is shown as a large pentagon.  This Cepheid also shows the largest mean magnitude difference relative to the KP and is flagged as having a nearby companion in \citet{freedman1994}.}
\end{figure}

\begin{figure}
\plotone{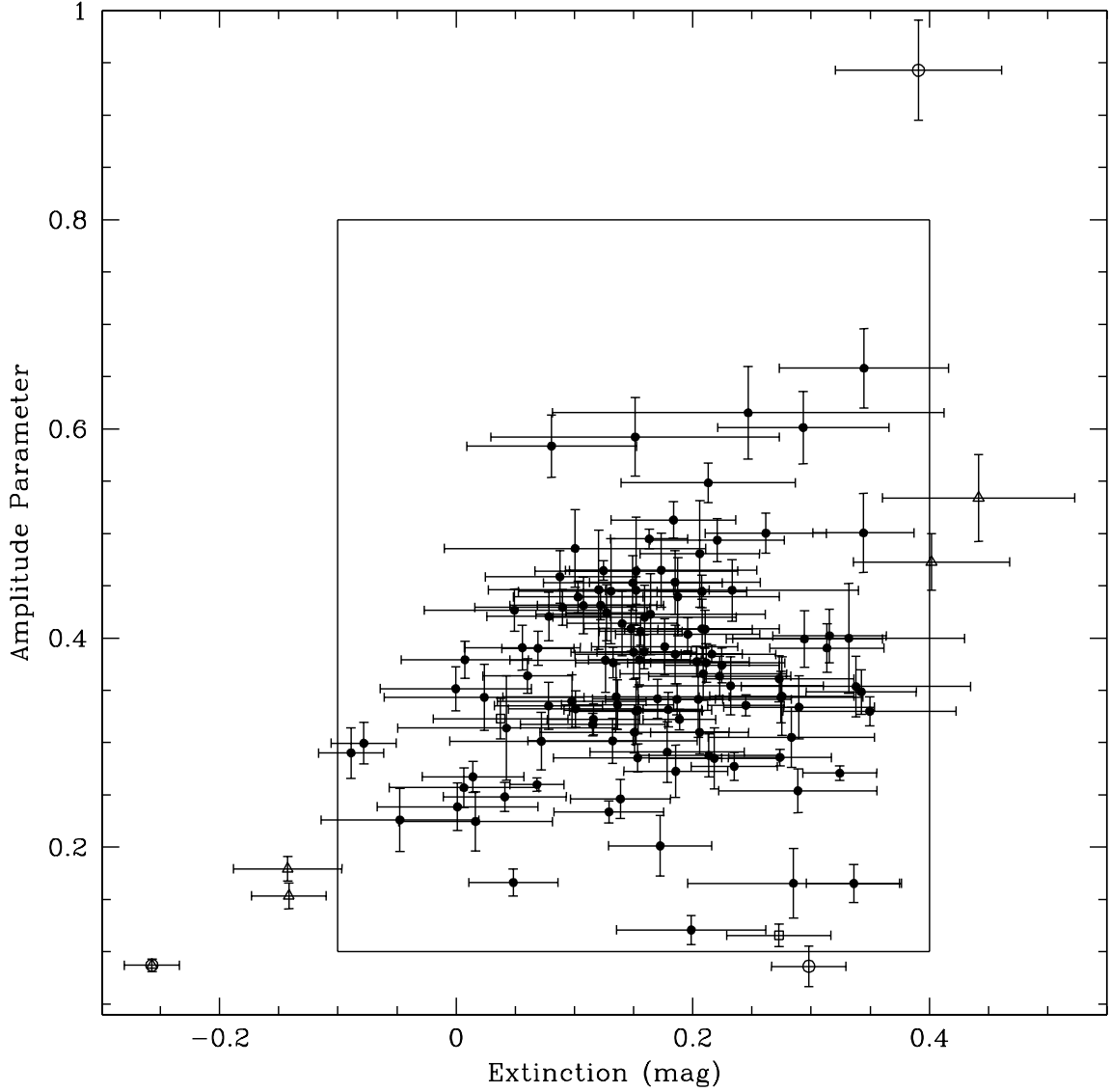}
\figurenum{3}
\label{fig:amp}
\caption{Amplitudes and extinctions for the 126 LBT Cepheids.  For our final analysis, we only use Cepheids within the  $-0.1\le E(\bv)\le0.4$ and $0.1\le A\le0.8$ selection box shown in the Figure. The Cepheids removed from the sample are marked according to the reason they were removed: extinction (triangles), crowding (squares), and amplitude (empty circles).}
\end{figure}

\begin{figure}
\plotone{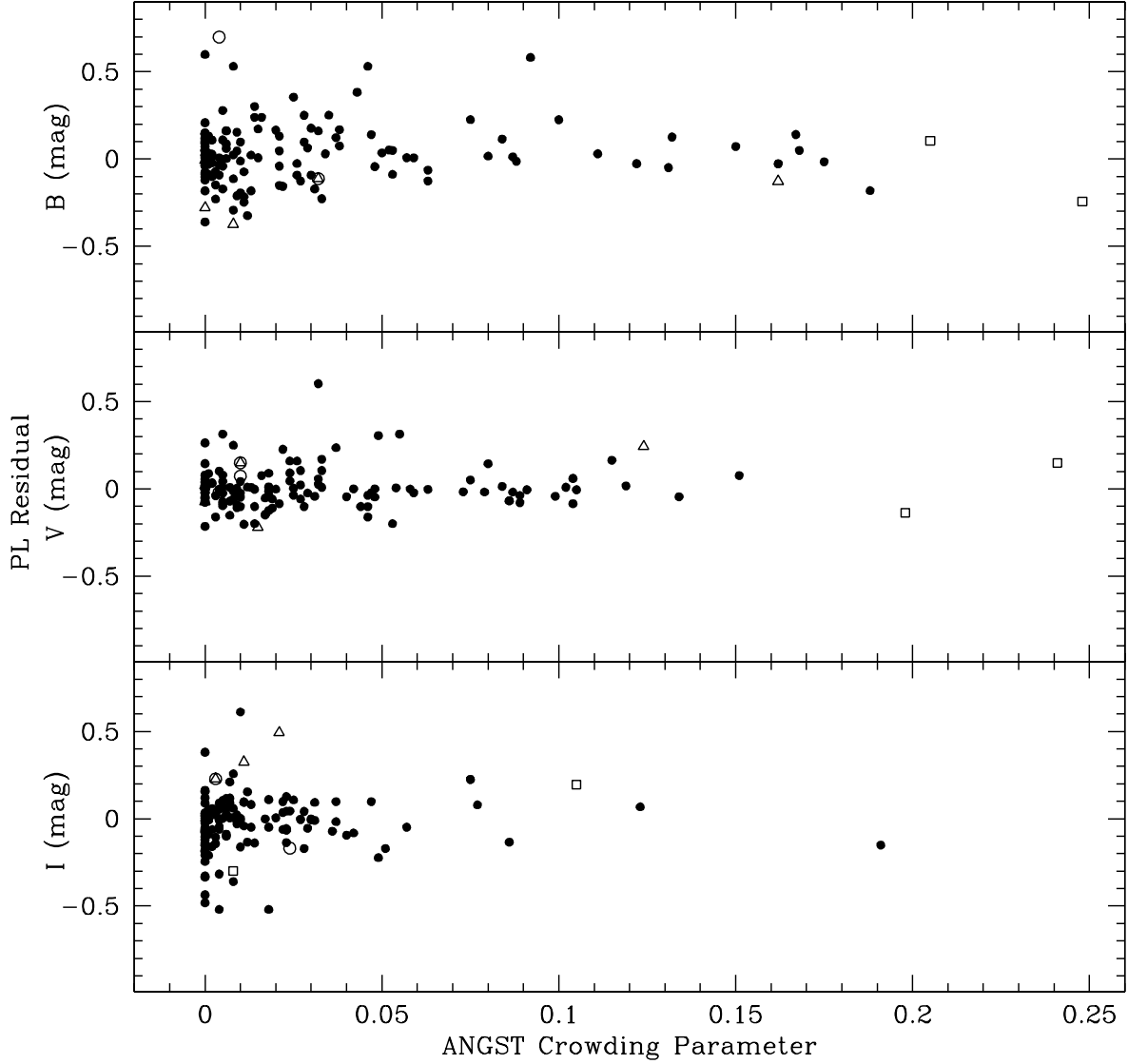}
\figurenum{4}
\label{fig:crowd}
\caption{ PL residuals as a function of the ANGST crowding parameter for the B,V, and I bands.  The filled circles represent the final sample.  There appear to be no correlations of residuals with crowding. As in Figure \ref{fig:amp}, the Cepheids removed from the sample are marked according to the reason they were removed: extinction (triangles), crowding (squares), and amplitude (empty circles). We restricted the final sample to have $ crowd \le 0.2$ in all filters and $(crowd_B + crowd_V + crowd_I) \le 0.4$ in total. }
\end{figure}

\begin{figure}
\plotone{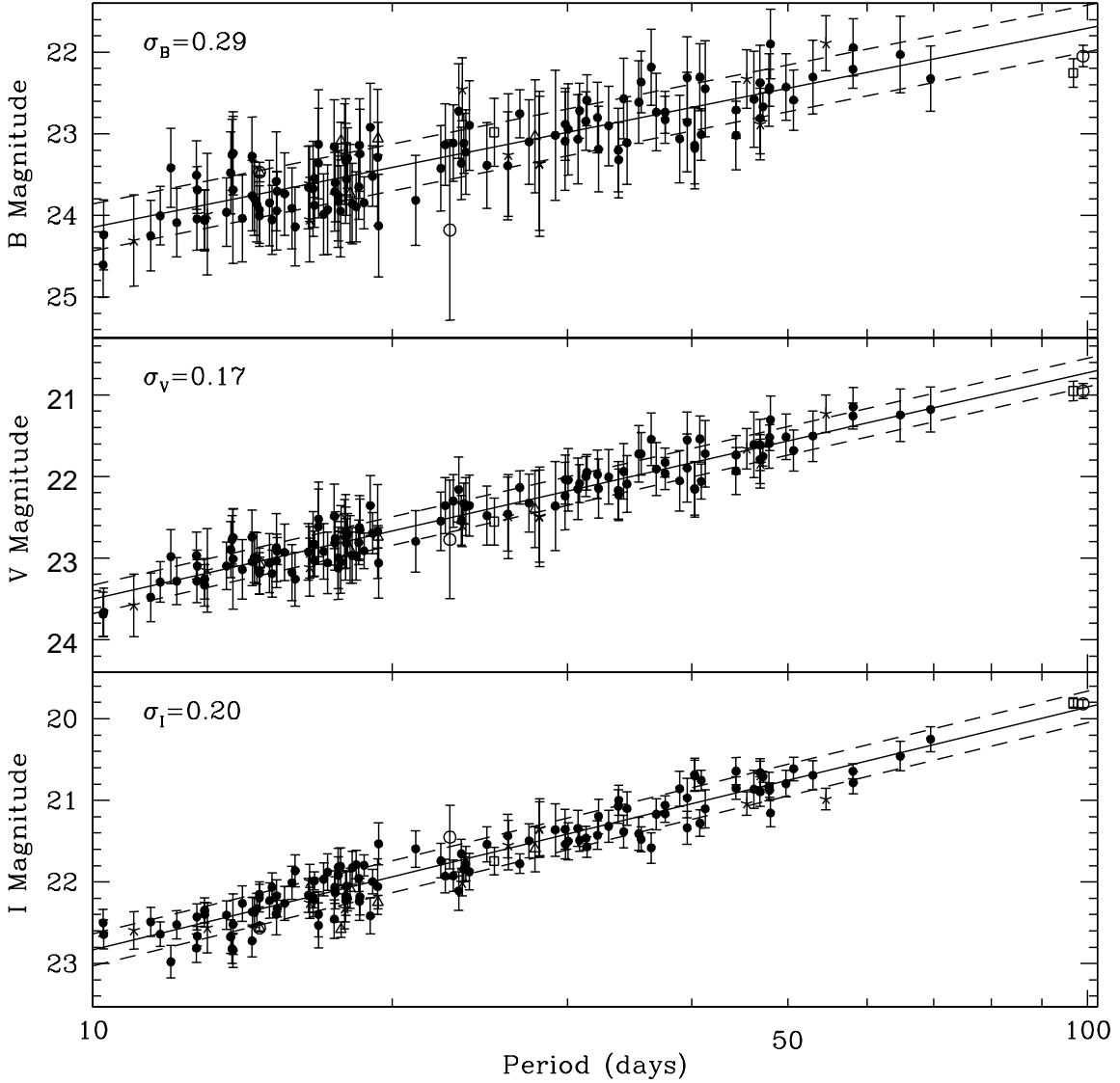}
\figurenum{5}
\label{fig:PL3rand}
\caption{The extinction-corrected random phase B, V, and I band PL relations.  The scaled ($\chi^2/{\rm dof}$=1) errorbars are shown and the Cepheid magnitudes have been corrected by the individually determined extinctions.  The solid lines show the OGLE PL relations for a relative distance modulus of $\Delta \mu_{LMC}=9.22\pm0.03$ mag and the dashed lines indicate the dispersions of the data about these relations.  The 107 Cepheids used for the final fit are filled circles.  Cepheids removed from the sample are also shown, marked as in the previous figures. Cepheids removed due to large residuals in the PL relations are shown with an asterisk.}
\end{figure}

\begin{figure}
\plotone{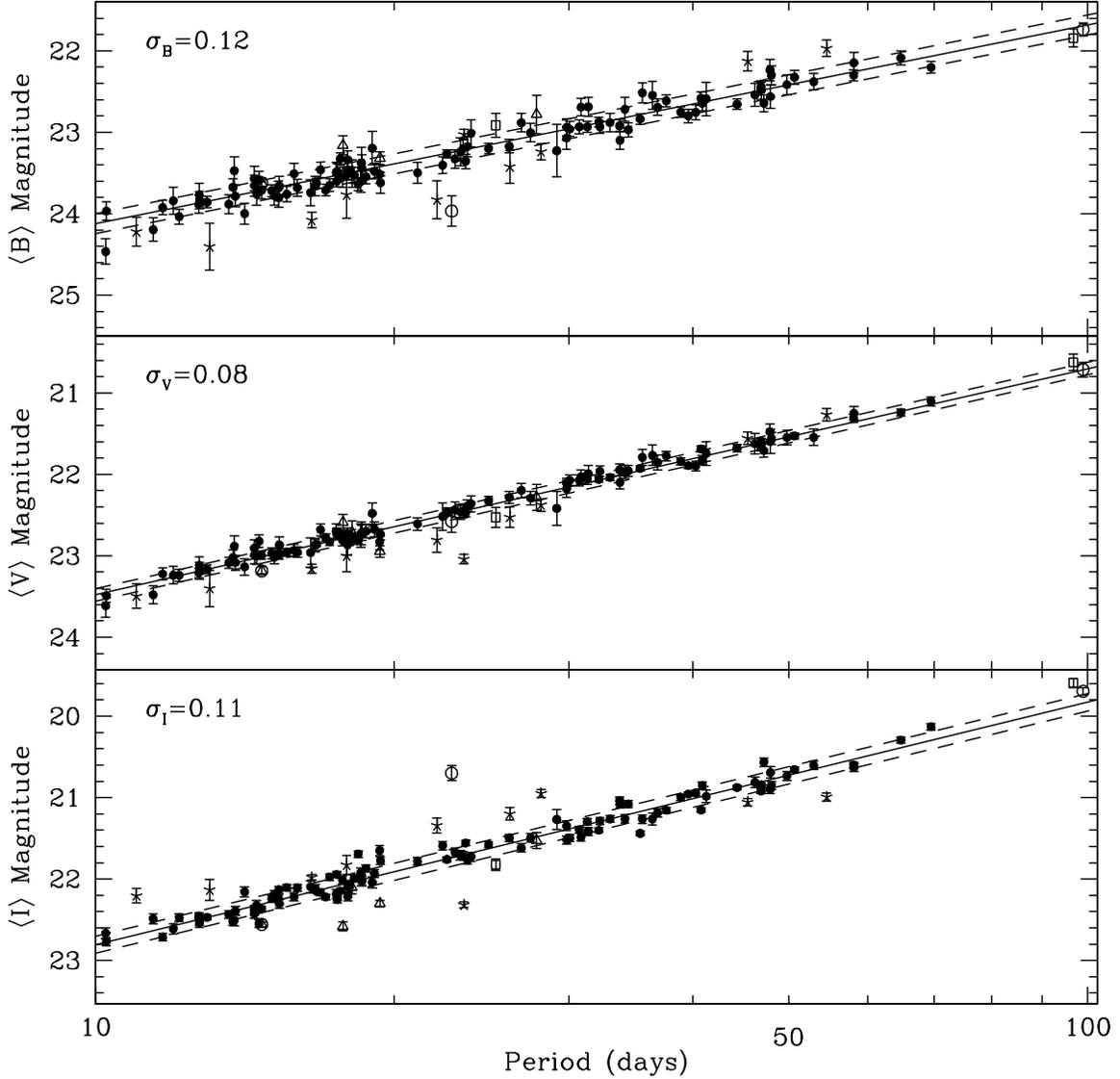}
\figurenum{6}
\label{fig:PL3}
\caption{The extinction-corrected, phase-averaged B, V, and I band PL relations.  The scaled errorbars are shown.  The solid lines show the OGLE PL relations for a relative distance modulus of $\Delta \mu_{LMC}=9.19\pm0.02$ mag, and the dashed lines indicate the dispersions of the data about these relations.  The 107 Cepheids used for the final fit are filled circles.  Cepheids removed from the sample are also shown, marked as in the previous figures. Cepheids removed due to large residuals in the PL relations are shown with an asterisk. }
\end{figure}

\begin{figure}
\plotone{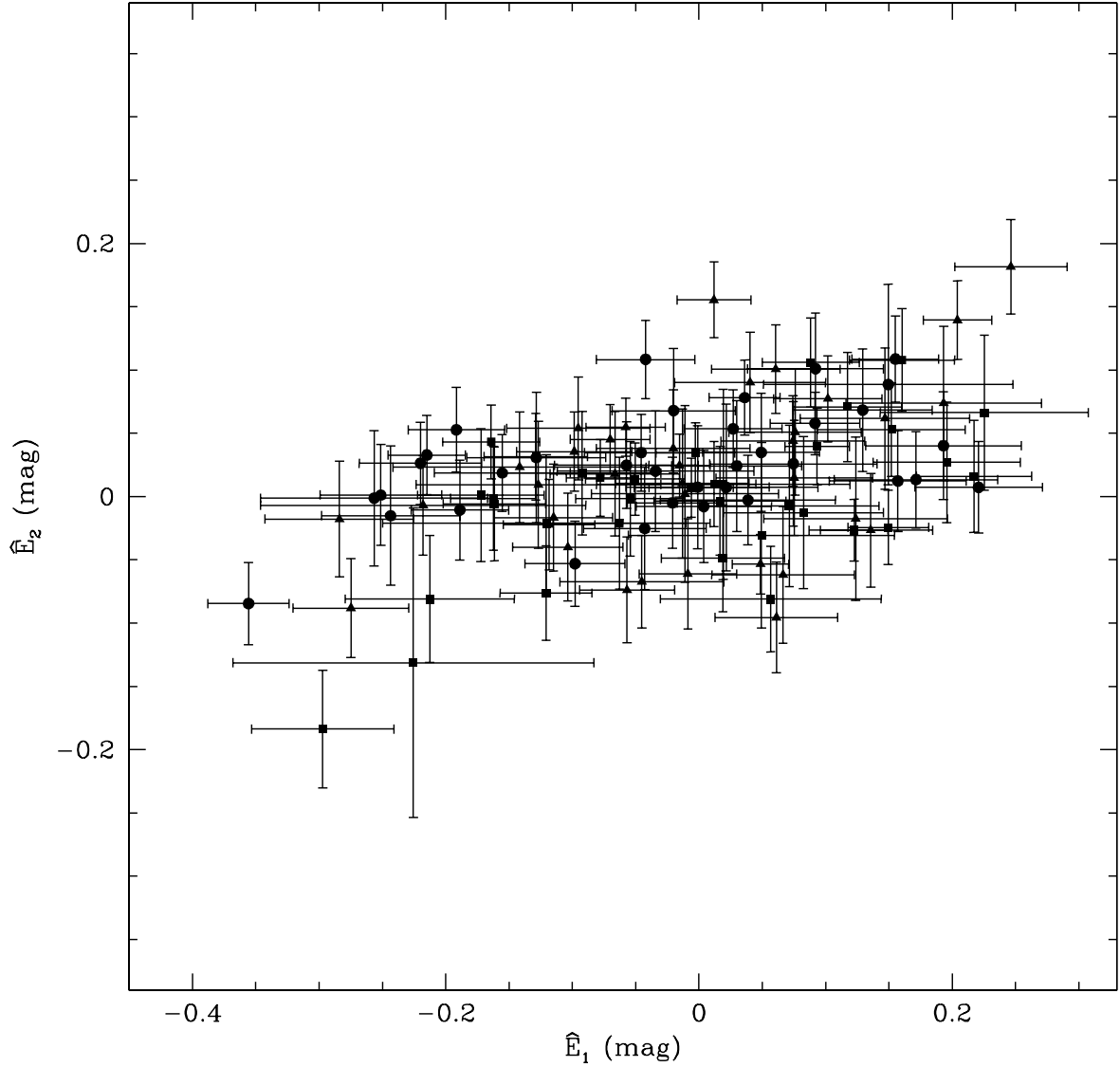}
\figurenum{7}
\label{fig:e1e2}
\caption{The $\hat{E}_1$ and $\hat{E}_2$ residuals for our final sample of 107 Cepheids, where $\vec{E}_1=\vec{\mu}-( \vec{\mu} \cdot \vec{R})\vec{R}/(\vec{R} \cdot \vec{R})$ corresponds to errors in distance that cannot be modeled as extinction and $\vec{E}_2= \vec{\mu} \times \vec{R}$, corresponds to residuals that can be modeled neither by changes in distance nor extinction.  We see a trend of $E_2$ residuals increasing with $E_1$ residuals, but the dispersion is dominated by $E_1$. The Cepheids in the inner-most radial bin marked by circles, those from the middle bin are triangles and the Cepheids in the outer-most bin are squares.}
\end{figure}

\begin{figure}
\plotone{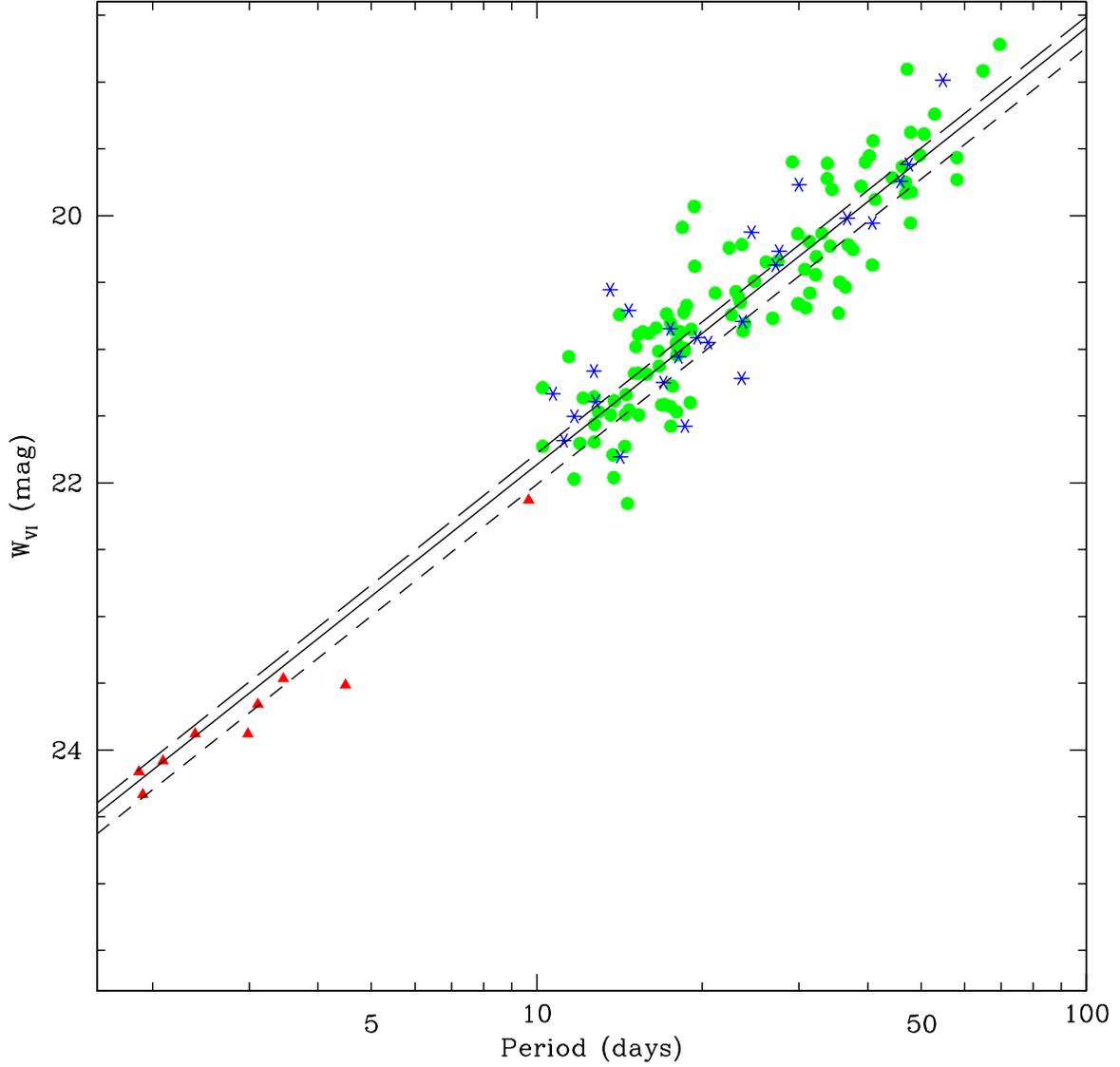}
\figurenum{8}
\label{fig:wmag}
\caption{The V/I band Wesenheit PL relation.  The green circles are the LBT Cepheids, the blue asterisks are the KP Cepheids and the red triangles are the \citet{mccommas2009} Cepheids.  The solid line is the PL for the LBT data, the long dashed line is the KP PL relation and the short dashed line is the PL from \citet{mccommas2009}.  For fits ignoring any metallicity dependence, our distance modulus falls between the previous results.}
\end{figure}

\begin{figure}
\plotone{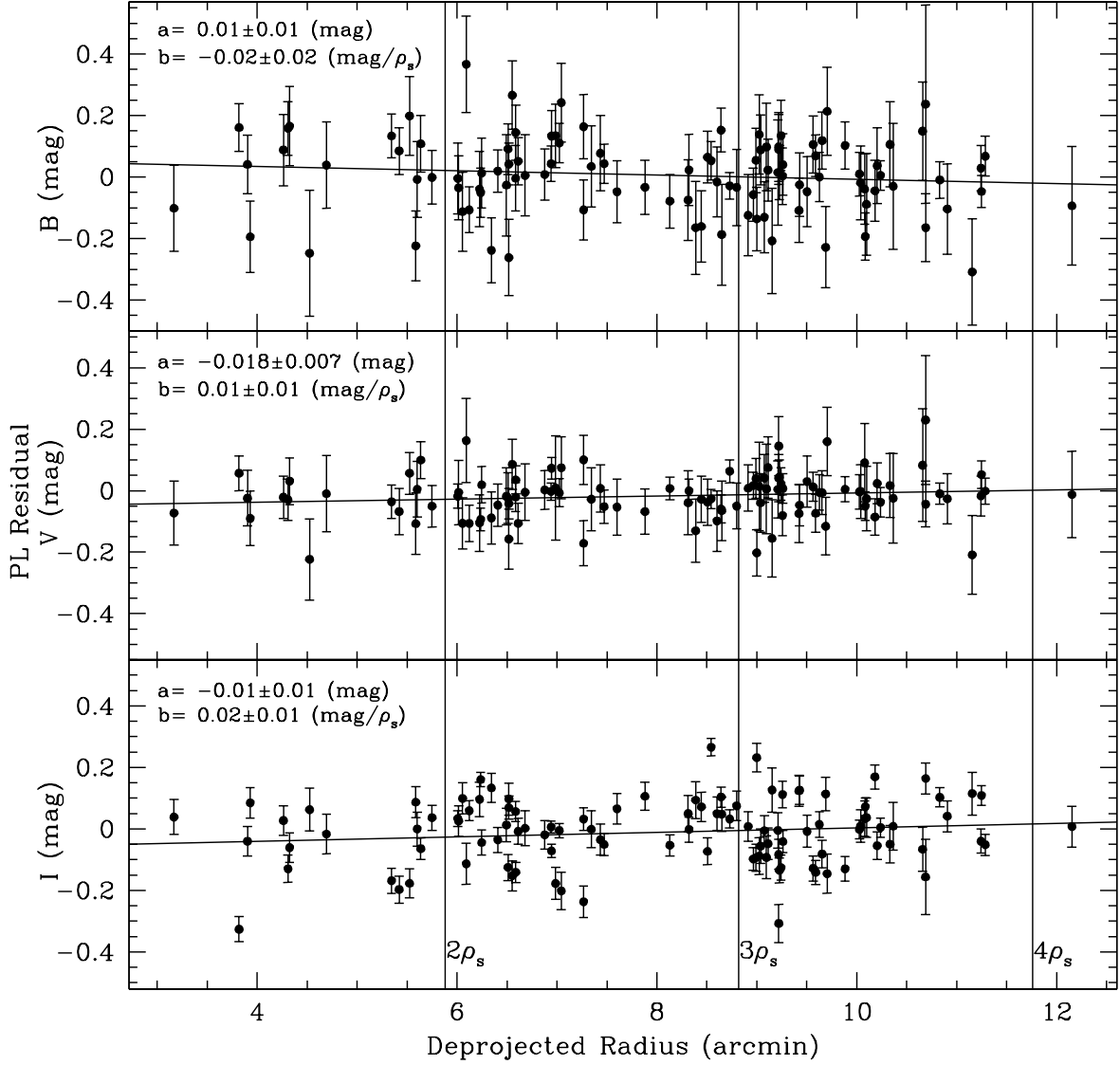}
\figurenum{9}
\label{fig:resid}
\caption{ PL residuals in the extinction-corrected mean magnitudes as a function of deprojected radius for the B, V, and I bands. The lines show linear fits to the residuals, a+b($\rho-2.7\rho_s$)$/\rho_s$, for each band where $2.7\rho_s$ is the mean radius.  We see a small negative slope for the B band, a slope close to zero in the V band, and a positive slope in the I band. The vertical lines mark several deprojected scale radii $\rho_s$.}
\end{figure}

\begin{figure}
\plotone{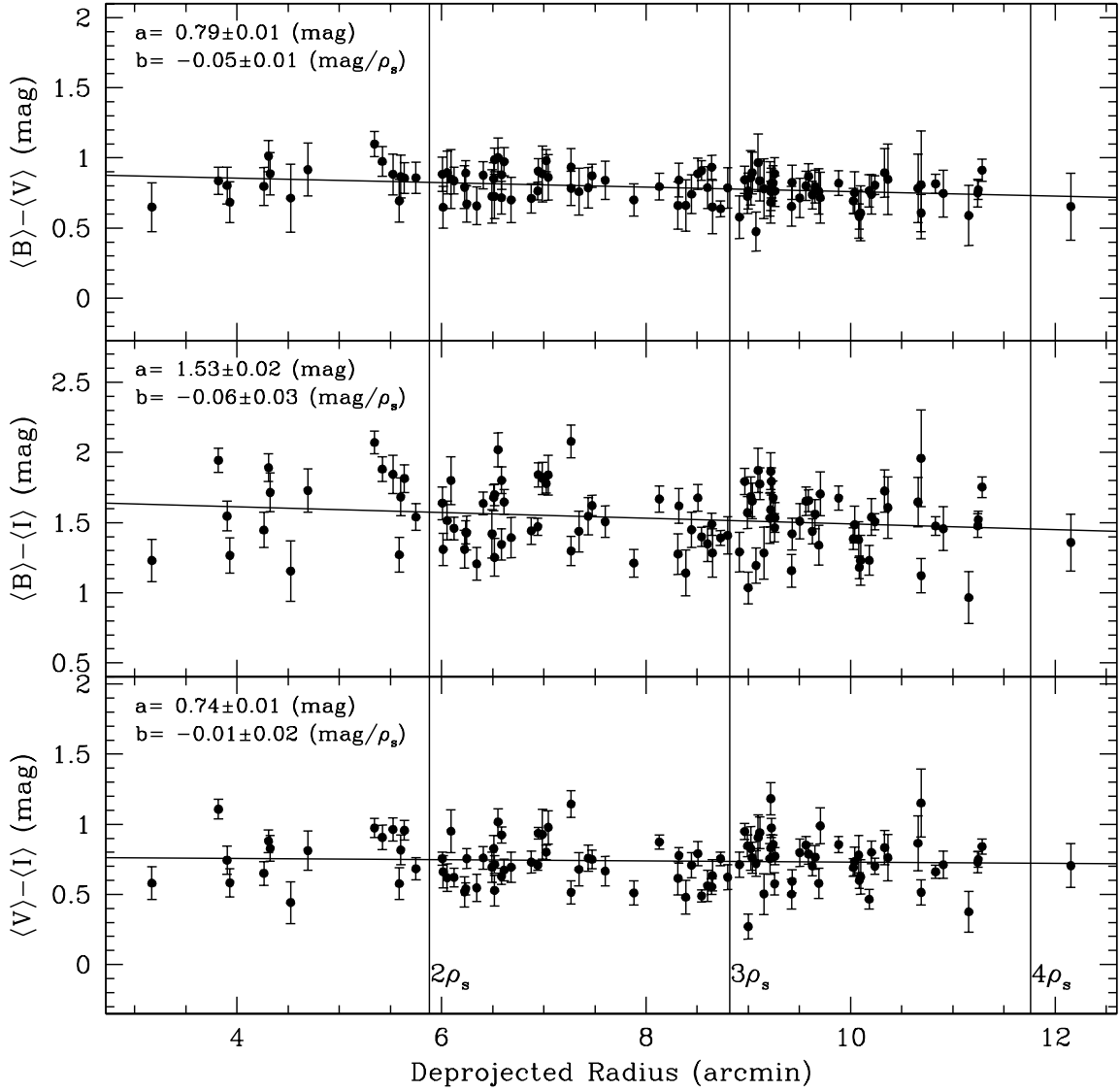}
\figurenum{10}
\label{fig:bvir}
\caption{Extinction-corrected colors as a function of deprojected radius.  Linear fits, a+b($\rho-2.7\rho_s$)$/\rho_s$, are again shown with solid lines for each color.  The vertical lines show several deprojected scale radii $\rho_s$. All Cepheids become bluer with increasing radius, as would be expected with decreasing metallicity. }
\end{figure}

\begin{figure}
\plotone{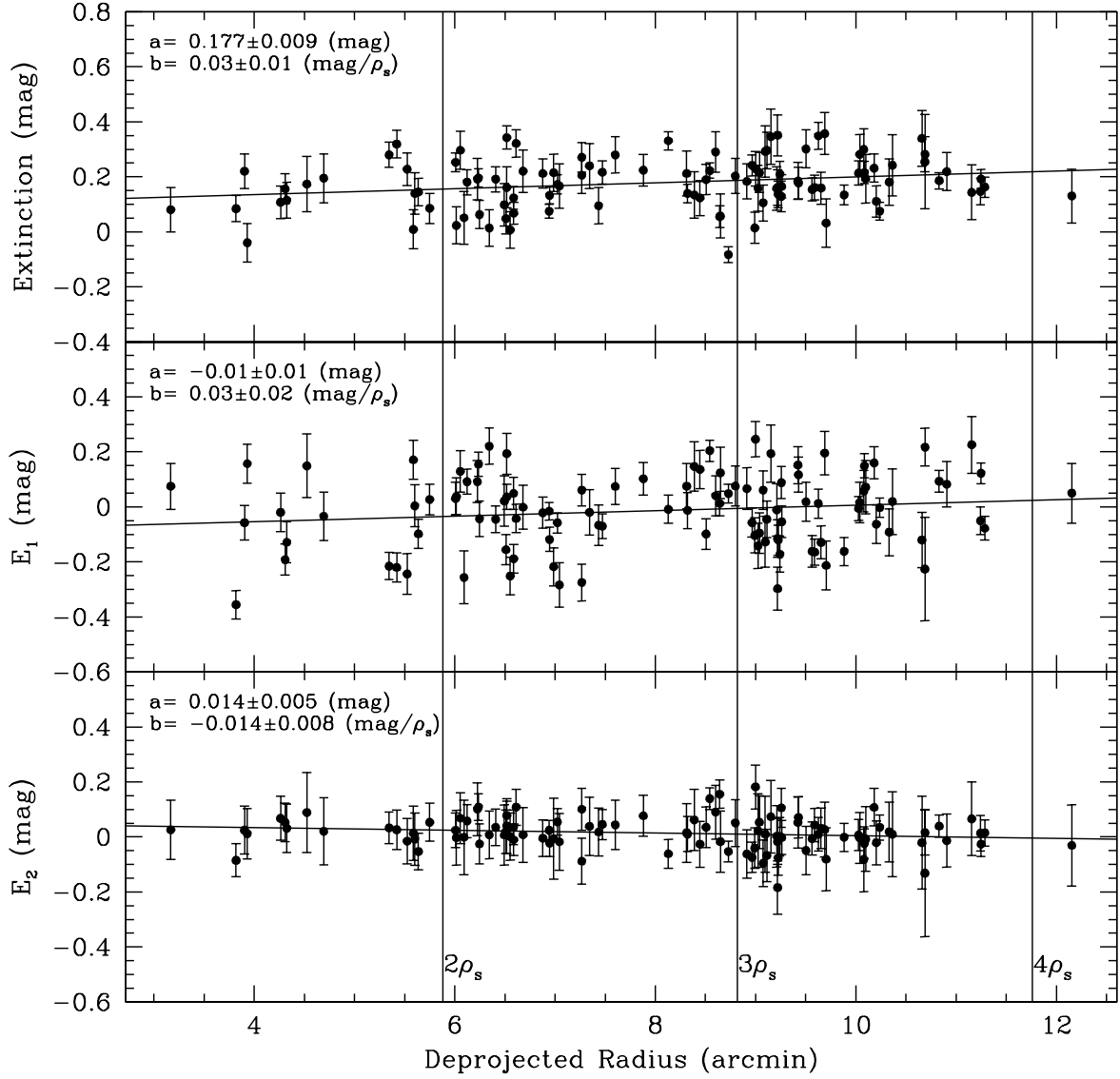}
\figurenum{11}
\label{fig:exr}
\caption{Extinction, $\hat{E_1}$ and  $\hat{E_2}$ residuals as a function of deprojected radius. The solid lines show a linear fit, a+b($\rho-2.7\rho_s$)$/\rho_s$, to the data in each panel.  The vertical lines show several deprojected scale radii $\rho_s$.
}
\end{figure}

\begin{figure}
\plotone{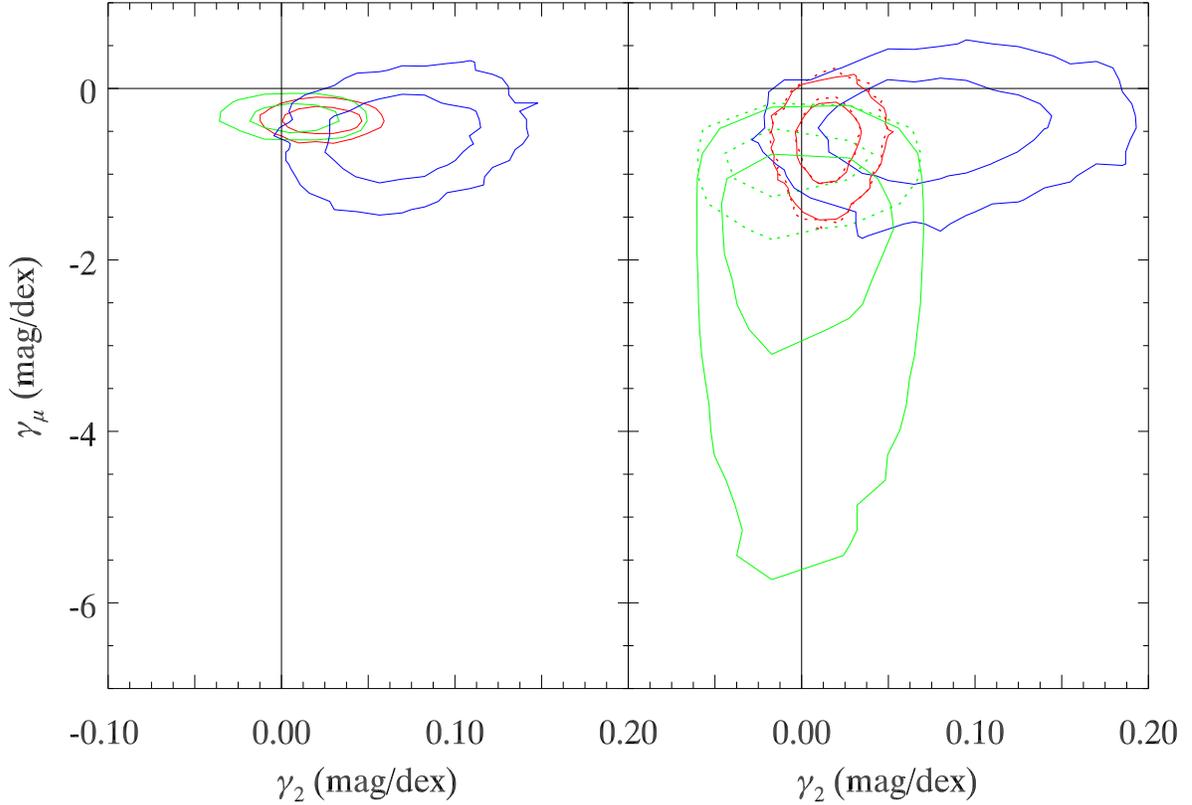}
\figurenum{12}
\label{fig:boot3}
\caption{The bootstrap uncertainty contours for the $\gamma_{\mu}$ and $\gamma_2$ metallicity correction parameters. The contours enclose 68.3\ and 95.4\ of the bootstrap estimates of the metallicity parameters.  The results from M81 data only are shown in blue, the NGC~4258 results are shown in green and the joint estimate is shown in red. The left panel fixes the abundance gradients to those in \citet{zaritsky1994}, while the right panel includes the fits to the \ion{H}{2} regions in the bootstrap procedure. In the right panel, the solid contours show the results using the \citet{zaritsky1994} data and the dotted contours include both the \citet{zaritsky1994} data and the shifted \citet{bresolin2011a} data.}
\end{figure}

\begin{figure}
\plotone{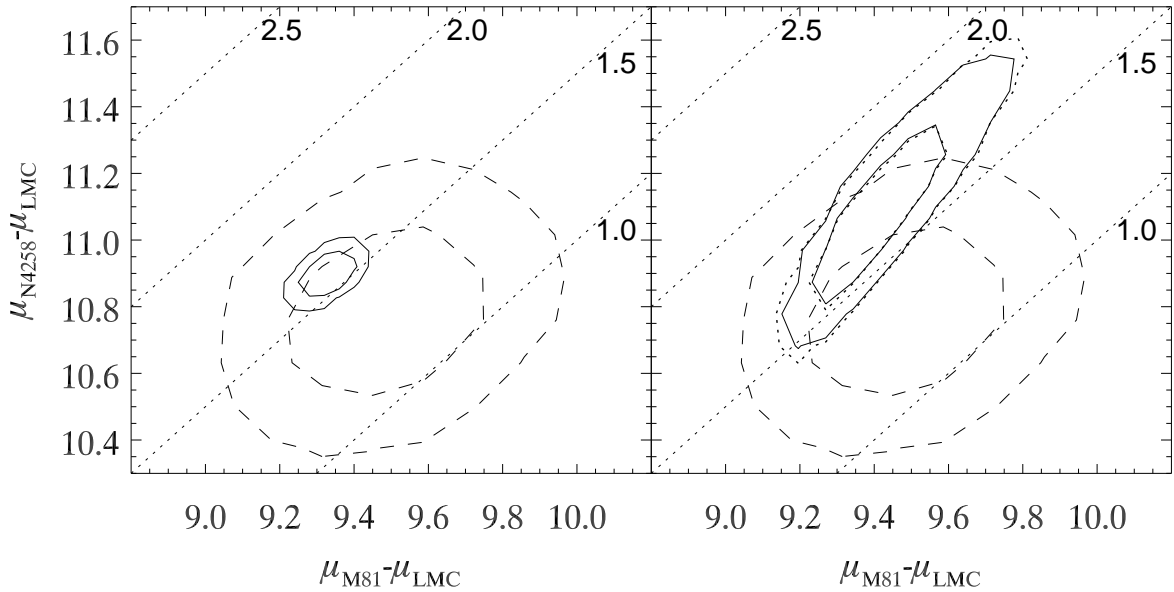}
\figurenum{13}
\label{fig:dist}
\caption{The uncertainties for $\mu_{N4258}-\mu_{LMC}$ as a function of $\mu_{M81}-\mu_{LMC}$. The solid contours are the 68.3\ and 95.4\ contours for the bootstrap distance estimates.  The left panel fixes the abundance gradients to those in \citet{zaritsky1994}, while the right panel includes the fits to the \ion{H}{2} regions in the bootstrap procedure.  In the right panel, the solid contours show the results using the \citet{zaritsky1994} data and the dotted contours include both the \citet{zaritsky1994} data and the shifted \citet{bresolin2011a} data.  Dotted lines of constant $\mu_{N4258}-\mu_{M81}$ are shown labeled with their value in magnitudes.  The dashed-line ellipses show the differential distances determined by geometric estimates.  We use distance moduli of $27.99\pm0.16$ mag for M81 \citep{bartel2007}, $29.29\pm0.15$ mag for NGC~4258 \citep{humph2008} and $18.5\pm0.1$ mag for the LMC \citep{freedman2001}.}
\end{figure}

\begin{figure}
\epsscale{.9}
\plotone{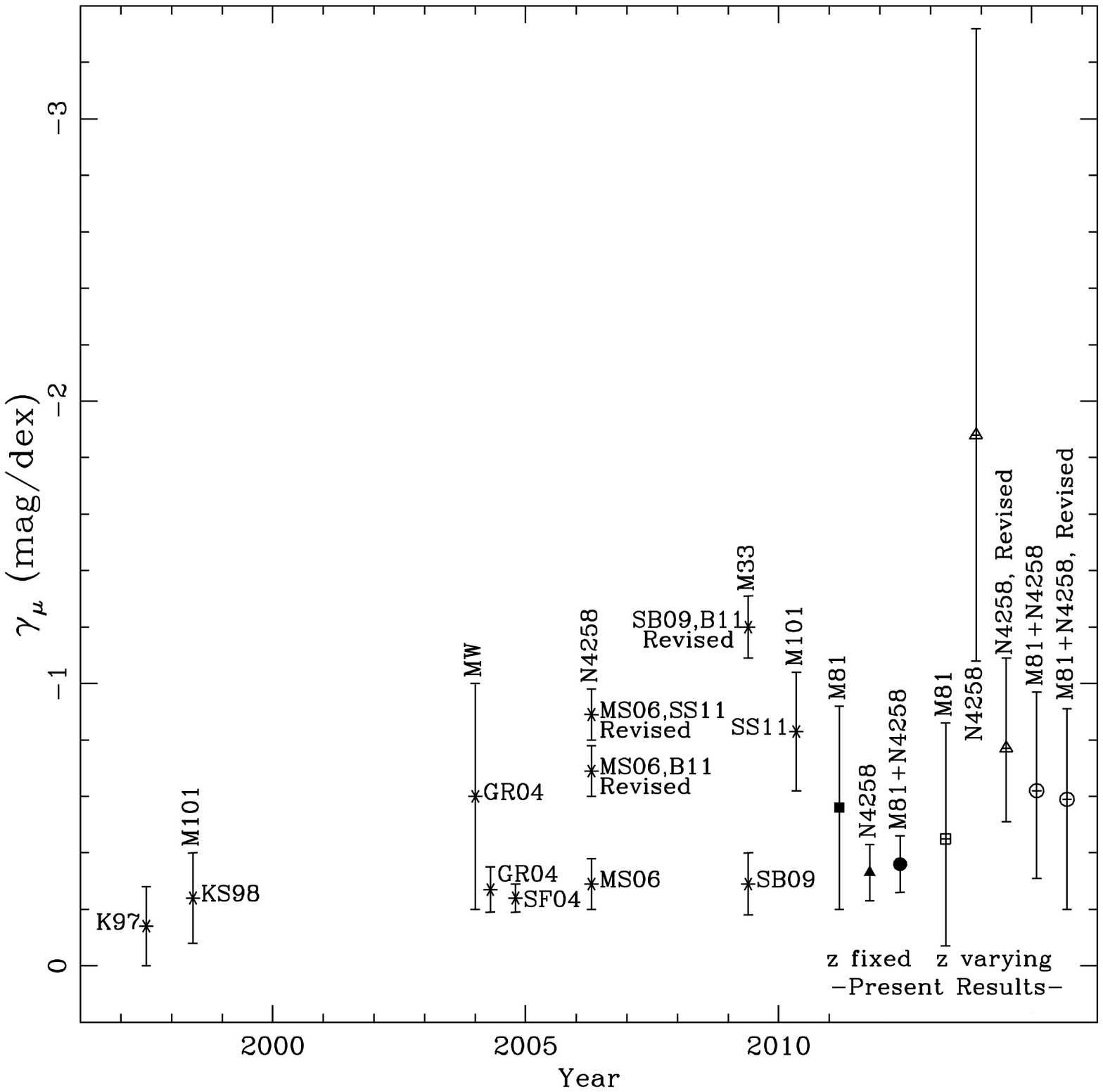}
\figurenum{14}
\label{fig:met}
\caption{Estimates of the Cepheid luminosity metallicity dependence, $\gamma_\mu$ in order of publication year.  The sources are  K97 (\citealt{kochanek1997}), KS98 (\citealt{kenni1998}), GR04 (\citealt{groen2004}), SF04 (\citealt{sakai2004}), MS06 (\citealt{macri2006}), SB09 (\citealt{scowcroft2009}), SS11 (\citealt{shappee2010}), B11 (\citealt{bresolin2011a}), and this work. The metallicity correction from \citet{macri2006} was revised in \citet{shappee2010} and \citet{bresolin2011a} based on different metallicity gradients. The ``Revised'' points for a reference show the metallicity corrections based on a revised metallicity gradient that were reported in the second source listed in the label.  If a study was done with a single galaxy, the result is also labeled with that galaxy.  The results from this work for the fixed metallicity gradients (``z fixed'') are marked with a solid symbols, while the results including the uncertainties of the metallicity gradients (``z varying'') are marked with open symbols.  The squares mark the results from M81 only, the triangles mark the results from NGC~4258 only, and the circles mark the joint results. The results from using a combination of \ion{H}{2} regions from \citet{zaritsky1994} and \citet{bresolin2011a} are labeled with ``Revised''.}
\end{figure}

\begin{deluxetable}{ccccccccc}
\tablecolumns{9}
\tablewidth{0pc}  
\tablecaption{ \label{tab:ceph} Cepheid Parameters} 
\tablehead{

  \colhead{ID} & \colhead{Period} &  \colhead{B} & \colhead{$\sigma_{B}$} &
  \colhead{V} & \colhead{$\sigma_{V}$} & \colhead{I} & \colhead{$\sigma_{I}$}
& \colhead{Flag}
}

\startdata

M81C~095614.95+690141.0	&	10.241	&	24.68	&	0.08	&	23.78	&	0.09	&	22.76	&	0.03	\\
M81C~095449.01+690118.1	&	10.256	&	24.42	&	0.07	&	23.84	&	0.03	&	22.97	&	0.03	\\
M81C~095612.27+690714.9	&	11.000	&	24.83	&	0.05	&	23.96	&	0.07	&	22.48	&	0.04	& OU\\
M81C~095633.24+685636.8	&	11.436	&	24.33	&	0.07	&	23.58	&	0.06	&	22.55	&	0.03	\\
M81C~095624.95+685843.6	&	11.690	&	24.88	&	0.03	&	23.96	&	0.04	&	23.15	&	0.03	\\
M81C~095538.72+685507.9	&	11.981	&	24.67	&	0.11	&	23.87	&	0.04	&	22.99	&	0.03	\\
M81C~095619.77+690646.0	&	12.146	&	24.71	&	0.04	&	23.76	&	0.03	&	22.78	&	0.02	\\
M81C~095520.88+690942.2	&	12.715	&	24.15	&	0.09	&	23.42	&	0.02	&	22.58	&	0.02	& C14 \\
M81C~095544.65+690527.4	&	12.724	&	24.11	&	0.08	&	23.38	&	0.06	&	22.69	&	0.02	\\
M81C~095536.75+690843.1	&	12.737	&	23.93	&	0.04	&	23.26	&	0.07	&	22.57	&	0.02	\\
M81C~095534.66+691213.7	&	12.954	&	24.77	&	0.03	&	23.87	&	0.03	&	22.89	&	0.01	& D \\
M81C~095449.34+690416.2	&	13.033	&	25.11	&	0.11	&	23.94	&	0.08	&	22.45	&	0.04	& OU \\
M81C~095533.10+690729.7	&	13.628	&	24.34	&	0.08	&	23.44	&	0.02	&	22.64	&	0.03	\\
M81C~095501.32+685901.8	&	13.760	&	24.47	&	0.004	&	23.63	&	0.06	&	22.88	&	0.03	\\
M81C~095536.27+691304.1	&	13.803	&	24.09	&	0.10	&	23.35	&	0.07	&	22.78	&	0.03	\\
M81C~095448.84+690512.9	&	13.836	&	24.73	&	0.08	&	23.81	&	0.04	&	22.82	&	0.03	& D \\
M81C~095608.62+690543.4	&	14.137	&	24.71	&	0.06	&	23.68	&	0.05	&	22.48	&	0.03	\\
M81C~095438.16+690928.7	&	14.459	&	24.14	&	0.09	&	23.35	&	0.05	&	22.69	&	0.03	\\
M81C~095439.57+690941.8	&	14.467	&	24.56	&	0.06	&	23.69	&	0.05	&	22.79	&	0.03	\\
M81C~095545.98+690904.2	&	14.537	&	24.79	&	0.06	&	23.79	&	0.05	&	22.79	&	0.03	\\
M81C~095620.85+685607.5	&	14.621	&	24.57	&	0.05	&	23.57	&	0.03	&	22.99	&	0.02	\\
M81C~095502.67+690954.4	&	14.707	&	24.24	&	0.09	&	23.40	&	0.02	&	22.61	&	0.02	& C9 D    \\
M81C~095601.51+690632.1	&	14.712	&	22.52	&	0.01	&	22.34	&	0.02	&	22.06	&	0.02	& EX AM\\
M81C~095625.06+690739.6	&	15.056	&	24.34	&	0.03	&	23.45	&	0.04	&	22.52	&	0.02	\\
M81C~095502.78+685902.1	&	15.146	&	24.47	&	0.06	&	23.55	&	0.03	&	22.50	&	0.03	\\
M81C~095613.28+685821.1	&	15.293	&	24.59	&	0.04	&	23.67	&	0.03	&	22.65	&	0.03	\\
M81C~095450.25+690053.1	&	15.303	&	24.71	&	0.09	&	23.68	&	0.02	&	22.54	&	0.03	\\
M81C~095441.33+691038.0	&	15.323	&	24.90	&	0.05	&	23.82	&	0.06	&	22.87	&	0.03	\\
M81C~095618.91+690649.1	&	15.593	&	24.42	&	0.07	&	23.46	&	0.01	&	22.40	&	0.03	& C31   \\
M81C~095505.31+691219.0	&	15.857	&	24.30	&	0.10	&	23.53	&	0.03	&	22.58	&	0.03	\\
M81C~095447.15+690145.4	&	15.976	&	23.74	&	0.07	&	23.00	&	0.02	&	22.14	&	0.03	\\
M81C~095424.12+691114.3	&	16.474	&	25.20	&	0.07	&	24.07	&	0.15	&	22.76	&	0.03	\\
M81C~095616.33+690344.7	&	16.522	&	24.79	&	0.06	&	23.71	&	0.01	&	22.30	&	0.01	& OU D\\
M81C~095615.22+690450.1	&	16.655	&	24.07	&	0.08	&	23.18	&	0.03	&	22.30	&	0.03	\\
M81C~095502.31+691017.2	&	16.703	&	23.94	&	0.05	&	23.10	&	0.01	&	22.30	&	0.01	& D \\
M81C~095617.78+690352.4	&	16.864	&	24.62	&	0.06	&	23.57	&	0.04	&	22.69	&	0.01	& D \\
M81C~095616.57+685615.1	&	17.066	&	23.94	&	0.04	&	22.96	&	0.02	&	22.33	&	0.02	\\
M81C~095621.41+690644.3	&	17.221	&	24.22	&	0.06	&	23.27	&	0.01	&	22.23	&	0.03	& C30  \\
M81C~095608.56+685846.0	&	17.492	&	24.31	&	0.06	&	23.33	&	0.04	&	22.55	&	0.03	\\
M81C~095441.10+690214.3	&	17.507	&	24.29	&	0.02	&	23.26	&	0.04	&	22.25	&	0.03	\\
M81C~095632.29+685538.7	&	17.528	&	24.47	&	0.07	&	23.48	&	0.02	&	22.70	&	0.02	\\
M81C~095623.69+690633.1	&	17.653	&	24.25	&	0.05	&	23.45	&	0.02	&	22.56	&	0.01	& D \\
M81C~095636.80+690332.8	&	17.745	&	24.02	&	0.09	&	23.17	&	0.02	&	22.23	&	0.03	\\
M81C~095528.70+690848.2	&	17.765	&	22.55	&	0.03	&	22.12	&	0.04	&	22.30	&	0.02	& EX \\
M81C~095408.28+690934.2	&	17.912	&	24.26	&	0.15	&	23.38	&	0.06	&	22.05	&	0.04	& OU\\
M81C~095430.81+690407.4	&	17.954	&	24.75	&	0.04	&	23.85	&	0.10	&	22.67	&	0.03	\\
M81C~095533.28+691252.3	&	17.964	&	24.55	&	0.06	&	23.66	&	0.03	&	22.76	&	0.03	\\
M81C~095616.19+690721.6	&	17.988	&	24.99	&	0.04	&	23.91	&	0.02	&	22.74	&	0.03	\\
M81C~095545.30+685559.5	&	18.027	&	23.11	&	0.01	&	22.56	&	0.02	&	21.92	&	0.02	\\
M81C~095628.20+690508.5	&	18.142	&	25.22	&	0.04	&	24.03	&	0.05	&	22.88	&	0.03	& EX \\
M81C~095455.40+690001.6	&	18.232	&	24.24	&	0.08	&	23.31	&	0.02	&	22.31	&	0.02	\\
M81C~095537.39+690648.8	&	18.402	&	24.00	&	0.04	&	23.08	&	0.02	&	21.86	&	0.03	\\
M81C~095442.63+690241.1	&	18.510	&	24.28	&	0.05	&	23.27	&	0.11	&	22.23	&	0.03	\\
M81C~095552.53+691237.2	&	18.537	&	23.93	&	0.13	&	23.15	&	0.09	&	22.27	&	0.02	\\
M81C~095614.42+690325.7	&	18.557	&	23.86	&	0.12	&	23.04	&	0.04	&	22.21	&	0.02	\\
M81C~095608.99+690457.0	&	18.738	&	23.75	&	0.03	&	22.85	&	0.06	&	21.96	&	0.02	\\
M81C~095506.04+690728.3	&	19.012	&	23.94	&	0.15	&	23.05	&	0.07	&	22.38	&	0.03	\\
M81C~095551.85+690505.8	&	19.107	&	24.42	&	0.04	&	23.40	&	0.06	&	22.36	&	0.03	\\
M81C~095635.80+690112.5	&	19.338	&	25.01	&	0.04	&	23.98	&	0.05	&	22.33	&	0.04	\\
M81C~095614.54+690707.0	&	19.351	&	22.71	&	0.02	&	22.48	&	0.03	&	22.03	&	0.02	& EX \\
M81C~095536.20+690820.7	&	19.377	&	24.59	&	0.10	&	23.49	&	0.02	&	22.22	&	0.03	\\
M81C~095538.07+690711.2	&	21.115	&	23.99	&	0.09	&	22.99	&	0.03	&	22.00	&	0.02	\\
M81C~095618.33+690145.0	&	22.088	&	24.54	&	0.13	&	23.35	&	0.04	&	21.66	&	0.03   & OU	\\
M81C~095621.06+690026.4	&	22.381	&	24.32	&	0.04	&	23.22	&	0.15	&	22.00	&	0.03	\\
M81C~095542.75+691149.2	&	22.604	&	23.58	&	0.02	&	22.70	&	0.03	&	21.90	&	0.02	\\
M81C~095610.64+685850.3	&	22.864	&	25.64	&	0.07	&	23.86	&	0.03	&	21.46	&	0.05	& AM \\
M81C~095623.11+690518.3	&	23.035	&	24.25	&	0.05	&	23.14	&	0.05	&	22.09	&	0.03	\\
M81C~095628.73+690409.2	&	23.325	&	23.91	&	0.13	&	22.98	&	0.04	&	22.01	&	0.03	\\
M81C~095516.05+691308.5	&	23.469	&	24.41	&	0.06	&	23.37	&	0.04	&	22.26	&	0.03	\\
M81C~095550.57+685918.8	&	23.521	&	22.73	&	0.01	&	22.82	&	0.04	&	22.18	&	0.02	& OU \\
M81C~095516.95+691005.3	&	23.610	&	23.65	&	0.07	&	22.70	&	0.01	&	21.69	&	0.02	& C12  \\
M81C~095620.34+690535.3	&	23.711	&	24.04	&	0.09	&	23.06	&	0.03	&	22.16	&	0.03	\\
M81C~095447.41+691124.7	&	23.911	&	23.26	&	0.12	&	22.55	&	0.04	&	21.84	&	0.02	\\
M81C~095502.25+690050.9	&	24.902	&	24.12	&	0.02	&	23.04	&	0.03	&	22.00	&	0.02	\\
M81C~095613.63+685929.4	&	25.330	&	23.08	&	0.03	&	22.65	&	0.06	&	21.90	&	0.02	& CR\\
M81C~095621.14+690500.0	&	26.132	&	23.98	&	0.03	&	22.91	&	0.05	&	21.86	&	0.03	\\
M81C~095631.27+685531.5	&	26.156	&	23.82	&	0.12	&	22.83	&	0.03	&	21.38	&	0.03	& OU\\
M81C~095552.35+690023.8	&	26.866	&	22.71	&	0.06	&	22.07	&	0.05	&	21.54	&	0.02	\\
M81C~095446.21+691223.7	&	27.456	&	24.30	&	0.06	&	23.28	&	0.04	&	22.08	&	0.03	\\
M81C~095624.57+690433.0	&	27.841	&	24.67	&	0.12	&	23.73	&	0.04	&	22.38	&	0.04	& EX \\
M81C~095506.00+690944.9	&	28.133	&	24.76	&	0.03	&	23.54	&	0.02	&	21.64	&	0.02	& OU D \\
M81C~095544.10+691205.7	&	29.168	&	24.32	&	0.21	&	23.25	&	0.10	&	21.76	&	0.05	\\
M81C~095423.75+691007.0	&	29.838	&	23.84	&	0.07	&	22.77	&	0.05	&	21.70	&	0.03	\\
M81C~095611.70+690751.9	&	29.857	&	23.70	&	0.06	&	22.69	&	0.02	&	21.86	&	0.03  & C28  \\
M81C~095614.61+690713.1	&	30.052	&	23.51	&	0.05	&	22.49	&	0.03	&	21.75	&	0.02  & C27   \\
M81C~095453.63+690627.6	&	30.738	&	23.30	&	0.04	&	22.35	&	0.03	&	21.56	&	0.02	\\
M81C~095538.64+690851.1	&	30.855	&	22.75	&	0.05	&	22.08	&	0.04	&	21.51	&	0.02	\\
M81C~095525.64+690940.9	&	31.312	&	23.75	&	0.02	&	22.68	&	0.05	&	21.67	&	0.03	\\
M81C~095610.21+690250.8	&	31.381	&	22.73	&	0.05	&	22.02	&	0.06	&	21.43	&	0.02	\\
M81C~095547.44+685406.8	&	32.164	&	23.67	&	0.04	&	22.67	&	0.01	&	21.76	&	0.02	\\
M81C~095543.97+690837.5	&	32.244	&	24.31	&	0.03	&	23.01	&	0.04	&	21.91	&	0.03	\\
M81C~095613.56+690620.0	&	33.009	&	23.48	&	0.11	&	22.50	&	0.01	&	21.53	&	0.03	& C29  \\
M81C~095527.07+690846.9	&	33.746	&	24.28	&	0.03	&	22.99	&	0.05	&	21.66	&	0.03	\\
M81C~095611.34+685836.6	&	33.777	&	23.13	&	0.06	&	22.12	&	0.04	&	21.10	&	0.02	\\
M81C~095518.79+691347.4	&	34.171	&	23.65	&	0.11	&	22.69	&	0.03	&	21.68	&	0.02	\\
M81C~095507.27+690717.5	&	34.435	&	23.64	&	0.04	&	22.47	&	0.03	&	21.38	&	0.03	\\
M81C~095444.83+690252.2	&	35.399	&	23.79	&	0.05	&	22.66	&	0.01	&	21.87	&	0.02	\\
M81C~095612.44+685849.5	&	35.585	&	23.21	&	0.07	&	22.32	&	0.05	&	21.58	&	0.02	\\
M81C~095500.25+685926.2	&	36.424	&	24.03	&	0.09	&	22.90	&	0.06	&	21.93	&	0.03	\\
M81C~095524.86+685808.7	&	36.859	&	23.89	&	0.04	&	22.77	&	0.06	&	21.73	&	0.03	\\
M81C~095542.18+690822.0	&	37.610	&	23.39	&	0.04	&	22.37	&	0.03	&	21.50	&	0.01	& D \\
M81C~095623.32+690800.2	&	38.902	&	23.45	&	0.04	&	22.37	&	0.01	&	21.31	&	0.02	\\
M81C~095613.84+690425.6	&	39.587	&	23.37	&	0.07	&	22.33	&	0.01	&	21.21	&	0.01	& D \\
M81C~095558.35+685845.0	&	40.282	&	23.37	&	0.06	&	22.37	&	0.03	&	21.22	&	0.01	& D \\
M81C~095506.63+690941.0	&	40.777	&	23.42	&	0.07	&	22.33	&	0.01	&	21.53	&	0.01 & 	C6   \\
M81C~095455.91+685959.2	&	40.890	&	23.23	&	0.06	&	22.27	&	0.02	&	21.12	&	0.03	\\
M81C~095525.63+691301.7	&	41.268	&	23.63	&	0.13	&	22.54	&	0.08	&	21.45	&	0.03	\\
M81C~095616.81+690342.8	&	44.326	&	23.40	&	0.04	&	22.24	&	0.03	&	21.21	&	0.01	& D \\
M81C~095621.72+690357.3	&	45.433	&	22.24	&	0.07	&	21.64	&	0.02	&	21.11	&	0.02	& OU \\
M81C~095502.66+690706.6	&	46.217	&	23.37	&	0.06	&	22.26	&	0.08	&	21.19	&	0.03	\\
M81C~095618.54+690843.1	&	46.866	&	23.26	&	0.02	&	22.30	&	0.03	&	21.29	&	0.02	\\
M81C~095509.57+690931.6	&	46.874	&	23.56	&	0.10	&	22.43	&	0.01	&	21.33	&	0.01	& C11   \\
M81C~095446.40+690441.6	&	47.176	&	23.52	&	0.05	&	22.38	&	0.04	&	20.96	&	0.03	\\
M81C~095527.14+691223.0	&	47.854	&	23.77	&	0.08	&	22.65	&	0.04	&	21.59	&	0.02	\\
M81C~095529.05+685633.9	&	47.896	&	23.81	&	0.06	&	22.55	&	0.11	&	21.26	&	0.04	\\
M81C~095515.49+685816.7	&	48.002	&	22.83	&	0.07	&	21.96	&	0.03	&	21.09	&	0.02	\\
M81C~095609.19+690319.8	&	49.759	&	23.01	&	0.06	&	22.01	&	0.04	&	21.00	&	0.03	\\
M81C~095525.41+685739.2	&	50.638	&	23.74	&	0.07	&	22.61	&	0.01	&	21.30	&	0.03	\\
M81C~095512.30+685801.2	&	52.954	&	23.65	&	0.04	&	22.52	&	0.07	&	21.18	&	0.03	\\
M81C~095633.84+690526.8	&	54.581	&	22.62	&	0.04	&	21.76	&	0.02	&	21.28	&	0.02	& OU \\
M81C~095545.91+690821.6	&	58.101	&	23.77	&	0.03	&	22.44	&	0.02	&	21.27	&	0.03	\\
M81C~095517.23+690936.2	&	58.136	&	23.41	&	0.08	&	22.22	&	0.04	&	21.21	&	0.02	\\
M81C~095610.62+690732.7	&	64.823	&	23.12	&	0.07	&	22.03	&	0.01	&	20.76	&	0.03	& C26  \\
M81C~095530.49+690833.2	&	69.541	&	23.40	&	0.03	&	22.02	&	0.02	&	20.67	&	0.03	\\
M81C~095611.68+685932.2	&	96.766	&	23.01	&	0.02	&	21.52	&	0.06	&	20.12	&	0.03	& CR \\
M81C~095621.16+690557.1	&	98.981	&	23.01	&	0.01	&	21.69	&	0.07	&	20.27	&	0.03	& AM \\
\enddata
\tablecomments{Cepheids removed from the final sample are flagged as follows. EX: extinction outside range; AM: amplitude outside range; CR: crowding parameter too large; OU: outlier in PL relations. We note the ID from \citet{freedman1994} for the Cepheids in common.  We also flag the Cepheids with two HST calibration observations with a 'D'. }

\end{deluxetable}

\begin{deluxetable}{cccccc}
\tablecolumns{6}
\tablewidth{0pc}  
\tablecaption{\label{tab:distnoz} Estimated Distance Moduli with No Metallicity Corrections.}
\tablehead{
  \colhead{Sample} & \colhead{PL source} & \colhead{Bands} & \colhead {Mean Magnitude} & \multicolumn{2}{c}{$\Delta\mu_{LMC}$}\\
                   &                     &                 &                           & \colhead{($\chi^2$)} & \colhead{(bootstrap)}}
\startdata

All & Updated &  BVI & Phase Averaged &$9.19\pm0.02$ & $9.19\pm0.03$\\
All & Updated &  BVI & Random Phase   &$9.22\pm0.03$ & $9.22\pm0.06$ \\

\tableline
%\\
All & Updated & VI & Phase Averaged   &$9.22\pm0.02$ & $9.22\pm0.04$ \\
%\\
All & Updated &  BV & Phase Averaged  &$9.07\pm0.02$ & $9.07\pm0.04$ \\
%\\
All & Updated & BI & Phase Averaged   &$9.17\pm0.02$ & $9.17\pm0.03$ \\
%\\
\tableline
%\\
All & Updated & VI & Random Phase     &$9.24\pm0.04$ & $9.24\pm0.07$ \\
%\\
All & Updated &  BV & Random Phase    &$9.10\pm0.03$ & $9.10\pm0.03$ \\
%\\
All & Updated & BI & Random Phase     &$9.21\pm0.04$ & $9.21\pm0.05$ \\
%\\
\tableline
%\\
All & OGLE99 & VI & Phase Averaged    &$9.25\pm0.02$ & $9.26\pm0.04$ \\
%\\
All & OGLE99 & VI & Random Phase      &$9.27\pm0.04$ & $9.27\pm0.07$ \\
%\\
\tableline
%\\
P $<$ 21.1 & Updated &  BVI & Phase Averaged &$9.18\pm0.02$ &$9.18\pm0.04$ \\
%\\
P $>$ 21.1 & Updated &  BVI & Phase Averaged &$9.20\pm0.02$ &$9.20\pm0.05$ \\
%\\
\tableline
%\\
$1.1\rho_s <$ R $\le 2.3\rho_s$ & Updated &  BVI & Phase Averaged &$9.20\pm0.03$ & $9.20\pm0.06$ \\
%\\
$2.4\rho_s \le$ R $\le 3.1\rho_s$ & Updated &  BVI & Phase Averaged &$9.17\pm0.03$ & $9.17\pm0.06$ \\
%\\
$3.1\rho_s <$ R $< 4.1\rho_s$ & Updated &  BVI & Phase Averaged &$9.21\pm0.03$ & $9.21\pm0.04$ \\
%\\
\tableline
%\\
$1.1\rho_s <$ R $\le 2.3\rho_s$ & Updated &  VI & Phase Averaged &$9.24\pm0.04$ & $9.24\pm0.08$ \\
%\\
$1.1\rho_s <$ R $\le 2.3\rho_s$ & Updated &  BV & Phase Averaged &$8.95\pm0.04$ & $8.95\pm0.05$ \\
%\\
$1.1\rho_s <$ R $\le 2.3\rho_s$ & Updated &  BI & Phase Averaged &$9.15\pm0.03$ & $9.15\pm0.05$ \\
%\\
\tableline
%\\
$2.4\rho_s \le$ R $\le 3.1\rho_s$ & Updated &  VI & Phase Averaged &$9.20\pm0.03$ & $9.20\pm0.07$ \\
%\\
$2.4\rho_s \le$ R $\le 3.1\rho_s$ & Updated &  BV & Phase Averaged &$9.09\pm0.04$ & $9.09\pm0.09$ \\
%\\
$2.4\rho_s \le$ R $\le 3.1\rho_s$ & Updated &  BI & Phase Averaged &$9.17\pm0.03$ & $9.17\pm0.04$ \\
%\\
\tableline
%\\
$3.1\rho_s <$ R $< 4.1\rho_s$ & Updated &  VI & Phase Averaged &$9.23\pm0.03$ & $9.23\pm0.05$ \\
%\\
$3.1\rho_s <$ R $< 4.1\rho_s$ & Updated &  BV & Phase Averaged &$9.16\pm0.04$ & $9.16\pm0.06$ \\
%\\
$3.1\rho_s <$ R $< 4.1\rho_s$ & Updated &  BI & Phase Averaged &$9.20\pm0.03$ & $9.20\pm0.04$ \\
%\\
\tableline
KP & OGLE99 &  VI & Averaged          &$9.25\pm0.08$ \\
ANGST & Updated &  VI & Phase Averaged &$9.34\pm0.05$ \\
\enddata
\tablecomments{OGLE99 refers to the original BVI PL relations in \citet{udalski1999}. Updated refers to the revised PL relations associated with the Cepheid catalogs. In this table the KP and ANGST distance moduli have had their metallicity corrections removed. }
\end{deluxetable}

\begin{deluxetable}{cccccc}
\tablecolumns{6}
\tablewidth{0pc}  
\tablecaption{\label{tab:distz}Estimated Distance Moduli with Metallicity Corrections.}
\tablehead{
  \colhead{Sample} & \colhead{PL source} & \colhead{Bands} & \colhead {Mean Magnitude} & \multicolumn{2}{c}{$\Delta\mu_{LMC}$}\\
                   &                     &                 &                           & \colhead{($\chi^2$)} & \colhead{(bootstrap)}}
\startdata
Final & Updated &  BVI & Phase Averaged &$9.39\pm0.08$ & $9.39\pm0.14$ \\
%\\
\tableline
%\\
$1.1\rho_s <$ R $\le 2.3\rho_s$ & Updated &  VI & Phase Averaged &$9.45\pm0.04$ & $9.44\pm0.08$  \\
%\\
$1.1\rho_s <$ R $\le 2.3\rho_s$ & Updated &  BV & Phase Averaged &$9.39\pm0.04$ & $9.39\pm0.05$ \\
%\\
$1.1\rho_s <$ R $\le 2.3\rho_s$ & Updated &  BI & Phase Averaged &$9.42\pm0.03$ & $9.42\pm0.05$ \\
%\\
\tableline
%\\
$2.3\rho_s \le$ R $\le 3.1\rho_s$ & Updated &  VI & Phase Averaged &$9.38\pm0.03$ & $9.37\pm0.07$  \\
%\\
$2.3\rho_s \le$ R $\le 3.1\rho_s$ & Updated &  BV & Phase Averaged &$9.45\pm0.04$ & $9.44\pm0.09$ \\
%\\
$2.3\rho_s \le$ R $\le 3.1\rho_s$ & Updated &  BI & Phase Averaged &$9.38\pm0.03$ & $9.38\pm0.04$  \\
%\\
\tableline
%\\
$3.1\rho_s <$ R $< 4.1\rho_s$ & Updated &  VI & Phase Averaged &$9.36\pm0.03$ & $9.36\pm0.05$ \\
%\\
$3.1\rho_s <$ R $< 4.1\rho_s$ & Updated &  BV & Phase Averaged &$9.44\pm0.04$ & $9.44\pm0.06$ \\
%\\
$3.1\rho_s <$ R $< 4.1\rho_s$ & Updated &  BI & Phase Averaged &$9.37\pm0.03$ & $9.37\pm0.04$ \\
%\\
\tableline
KP & OGLE99 &  VI & Averaged          &$9.30\pm0.08$ \\
ANGST & Updated &  VI & Phase Averaged &$9.37\pm0.05$ \\
\enddata
\tablecomments{OGLE99 refers to the original BVI PL relations in \citet{udalski1999}. Updated refers to the revised PL relations associated with the Cepheid catalogs. In this table the KP and ANGST distance moduli have been corrected for metallicity. }
\end{deluxetable}

\begin{deluxetable}{ccc}
\tablecolumns{3}c
\tablewidth{0pc}  
\tablecaption{\label{tab:distcomp}Comparison with Other Distance Methods.}
\tablehead{
  \colhead{Source} & \colhead{Method} & \colhead{Distance Modulus}  }
 \startdata
 M81 only, fixed gradient  & Cepheid & $27.80\pm0.14$ \\
 M81+N4258, fixed gradient  & Cepheid & $27.74\pm0.05$ \\
 M81+N4258, uncertain gradient  & Cepheid & $27.81^{+0.15}_{-0.11}$ \\
 M81+N4258, updated uncertain gradient  & Cepheid & $27.80^{+0.11}_{-0.16}$ \\
\citet{sakai2004} & TRGB & $28.03\pm0.12$ \\
\citet{jensen2003} & SBF & $27.71\pm0.26$ \\
\citet{bartel2007} & ESM & $27.99\pm0.16$ \\

\enddata

\tablecomments{ Calibrated to an LMC distance modulus of 18.41 mag.  TRGB: tip of the red giant branch, SBF: surface brightness fluctuations, ESM: expanding shock method. }
\end{deluxetable}

\end{document}